\documentclass[useAMS,usenatbib]{mn2e}
\usepackage{subfigure}
\usepackage{graphicx}
\usepackage{color}
\usepackage{soul}
\title[Black Hole Binaries]{Compact Binaries in Star Clusters II - Escapers and Detection Rates}
\author[J. M. B. Downing, M. J. Benacquista, M. Giersz, \& R. Spurzem]{J. M. B. Downing$^{1,2}$\thanks{E-mail: downin@ari.uni-heidelberg.de}, M. J. Benacquista$^{3}$, M. Giersz$^{4}$, and R. Spurzem$^{5,6,1}$\\
  $^{1}$Astronomisches Rechen-Institut, Zentrum f\"{u}r Astronomie der
  Universit\"{a}t Heidelberg, Monchhofsra\ss e 12-14,\\ D-69120 Heidelberg,
  Germany\\
  $^{2}$Fellow of the International Max-Planck Research School for Astronomy
  and Cosmic Physics at the University of Heidelberg,\\ Heidelberg, Germany\\
  $^{3}$Center for Gravitational Wave Astronomy, University of Texas at
  Brownsville, Brownsville, TX 78520, USA\\
  $^{4}$Nicolaus Copernicus Astronomical Center, Polish Academy of Sciences,
  ul. Bartycka 18, 00-716 Warsaw, Poland\\
  $^{5}$National Astronomical Observatories, Chinese Academy of Sciences,
  20A Datun Rd., Chaoyang District, 100012, China\\
  $^{6}$Kavli Institute of Astronomy and Astrophysics, Peking University,
  Beijing, China}
\begin{document}
\date{Accepted ... Received ... in original form ...}
\pagerange{\pageref{firstpage}--\pageref{lastpage}} \pubyear{2010}
\maketitle
\label{firstpage}
\begin{abstract}

We use a self-consistent Monte Carlo treatment of stellar dynamics to investigate black hole binaries that are dynamically ejected from globular clusters to determine if they will be gravitational wave sources.  We find that many of the ejected binaries have initially short periods and will merge within a Hubble time due to gravitational wave radiation.  Thus they are potential sources for ground-based gravitational wave detectors.  We estimate the yearly detection rate for current and advanced ground-based detectors and find a modest enhancement over the rate predicted for binaries produced by pure stellar evolution in galactic fields.  We also find that many of the ejected binaries will pass through the longer wavelength Laser Interferometer Space Antenna (LISA) band and may be individually resolvable.  We find a low probability that the Galaxy will contain a binary in the LISA band during its three-year mission.  Some such binaries may, however, be detectable at Mpc distances implying that there may be resolvable stellar-mass LISA sources beyond our Galaxy.  We conclude that globular clusters have a significant effect on the detection rate of ground-based detectors and may produce interesting LISA sources in local group galaxies.

\end{abstract}

\section{Introduction}
\label{sec:intro}

Gravitational waves, metric perturbations in space-time caused by time-varying mass-energy distributions, offer a new window on the universe that is independent of the electromagnetic spectrum.  In particular the inspiral and merger of stellar-mass compact binaries (binaries consisting of neutron stars (NSs) and/or black holes (BHs)) are predicted to be major burst sources for the high-frequency ground-based gravitational wave detectors Virgo and LIGO.  The current generation of these detectors should be able to detect such events out the the Virgo cluster \citep{Abbott05,Abbott06,Abbott10} while upgrades to these detectors (e.g. advanced LIGO) should lead to detections at cosmologically significant distances.  At larger separations compact binaries, including white dwarf (WD) binaries, within the Galaxy may be detected by the low-frequency space-based gravitational wave detector LISA planned for 2018-2020 \citep{Hils90,Benacquista01,Nelemans01,BBandB08}.  WD-WD binaries are expected to be plentiful and will manifest as a confusion limited noise source \citep{Evans87,Hils90,Nelemans01,Timpano06,Ruiter10} whereas the rarer NS-NS, NS-BH and BH-BH binaries may be individually resolved.

Many of the sources producing gravitational waves will not emit significant amounts of electromagnetic radiation (particularly NS-NS and BH-BH binaries) and it is necessary to perform population synthesis models in order to constrain event rates and produce templates for gravitational wave detectors.  Significant efforts have been made to constrain the population of compact binaries in galactic fields (e.g. \citealt{SPZandY98,FWandH99,BKandB02,Belczynski07}) and these studies have found that detection rates should be dominated by NS-NS inspirals.  Although BH-BH
binaries are more massive and can be detected with a higher signal-to-noise ratio at larger distances, NSs are both more plentiful due to the power-law
shape of the initial mass function (IMF) and the progenitors of NS-NS binaries are less likely to undergo mass-transfer leading to a merger than are their
BH-BH counterparts.  In particular \cite{Belczynski07} determined advanced LIGO detection rates of $\sim 20$ yr$^{-1}$ for NS-NS binaries, $\sim 2$
yr$^{-1}$ for BH-BH binaries and only $\sim 1$ yr$^{-1}$ for NS-BH binaries.  \cite{BBandB08} have similarly concluded that the number of resolved
stellar-mass detections by LISA, although small, will also be dominated by NS-NS binaries.

Although BH-BH binaries may be rare in galactic fields, it is possible to produce such binaries through dynamical interactions in star clusters \citep{SigPhin93} and thus enhance the total number of BH-BH mergers in the universe.  There are two types of interaction that create BH-BH binaries.  The first is  few-body binary formation where a close encounter between multiple stars transfers kinetic energy to one star at the expense of the relative energy between two others.  This leaves a bound pair and an escaper.  The second is exchange where a single star interacts with a binary and is exchanged into the binary  in place of one of the original members.  Due to equipartition of energy both of these processes favour the escape of the lightest member of the interaction \citep{HandF80,HHandM96} and thus preferentially introduce massive objects, such as BHs, into binary systems.  Another important effect is binary hardening where ``hard'' binaries (binaries with a binding energy greater than the kinetic thermal energy of the cluster) can have their binding energy increased and thus  their period shortened \citep{Heggie75}.  This process reduces the separation between the binary members and thus can bring a BH-BH binary into the gravitation wave regime more quickly than would isolated evolution in galactic fields.  Interactions between binaries can result in both exchange and  binary hardening for both binaries.  Finally interactions can destroy binaries if the kinetic energies of the centres of mass are sufficiently high and thus BH-BH binaries can also be removed by dynamical interactions.  All of these processes are enhanced by high stellar densities and thus proceed most rapidly in the very concentrated cores of star clusters.  Equipartition of energy requires that the most massive objects in a star cluster sink to the centre in a process called mass-segregation \citep{Spitzer87}.  Since BHs and binaries rapidly become the most massive objects in the system they will rapidly migrate to the centre of the cluster where interaction rates are highest.  This means that BHs will tend to experience a disproportionately large number of interactions and this further enhances the effect of star cluster dynamics on the BH-BH binary population.

The effect of star cluster dynamics have been studied before using limited dynamical models.  \cite{GMandH04} and \cite{OLeary06} have investigated the BH-BH binary population in star clusters assuming the BHs form completely mass-segregated subsystem that interacts only with itself.  They find that the BHs and BH-BH binaries interact strongly with each other leading to both the formation and destruction of BH-BH binaries.  Depending on the initial conditions  assumed for their clusters, \cite{OLeary06} find $\sim 1-10$ BH-BH merger detections per year using the parameters for advanced LIGO.  Up to $70\%$ of these  mergers occur in BH-BH binaries that have been dynamically ejected from the clusters.  These simulations contained no treatment of stellar evolution.  By contrast \cite{Sadowski08} have performed cluster simulations with stellar evolution but assuming that the BHs and BH-BH binaries always remain in dynamical equilibrium with the rest of the cluster.  This is in some sense the opposite dynamical assumption made in \cite{OLeary06} and leads to a much  higher merger rate of $\sim 25-3000$ detections per year, depending again on the initial conditions.  This is because although there are fewer interactions that form BH-BH binaries, once formed such a binary is unlikely to interact with another BH and be disrupted.  They claim only $\sim 10\%$ of mergers occurred outside the clusters.  Finally \cite{SPZandM00} have performed direct $N$-body simulations of star clusters focusing on BHs and the BH-BH binary population.  These simulations include full cluster dynamics but contain only $\sim 4096$ particles and as such are much smaller than real globular
clusters.  \cite{SPZandM00} find a detection rate of a few detections per year to almost one a day depending on the assumed globular cluster formation
history and the composition of the globular cluster population.  They also find that the vast majority of the BH-BH binaries are ejected from the cluster and thus most mergers occur in the field.  Consequently their results are more consistent with those of \cite{OLeary06} than \cite{Sadowski08}.

In \cite{Downing10} (hereafter Paper I) we have revisited this problem using a Monte Carlo star cluster simulation code that includes stellar evolution, a
fully self-consistent treatment of the global cluster dynamics, and is capable of simulating a number of particles comparable to real globular clusters.  We
demonstrate that the BHs strongly mass-segregate and thus confirm that the approximation made in \cite{OLeary06} is favoured over that of \cite{Sadowski08}.  We found several potential LISA sources in our simulations, four of which would be detectable in nearby globular clusters.  Our results differ from \cite{OLeary06} in that we find no BH-BH mergers within our clusters, possibly due to our more approximate treatment of few-body interactions.  This result is, however, consistent with the direct $N$-body results of \cite{SPZandM00}.  In Paper I we found that many BH-BH binaries were ejected from the cores of our clusters with very high binding energies, consistent with the results of both \cite{SPZandM00} and \cite{OLeary06}.  These binaries are massive enough and have short enough periods that they should merge in a galactic field within a Hubble time ($T_{H}$).  Investigating these escaping binaries is the subject of this paper.  In \S~\ref{sec:methods} we briefly describe our Monte Carlo code and initial conditions.  In \S~\ref{sec:properties} we describe the properties of the escaping binaries.  In \S~\ref{sec:mergers} we consider the merger rate of escaping binaries and we translate this into a detection rate for ground-based gravitational wave detectors in \S~\ref{sec:LIGOdet}.  In \S~\ref{sec:LISAdet} we consider whether any of our binaries can be detected by the planned LISA space-based detector.  We discuss our results in \S~\ref{sec:discussion} and conclude in \S~\ref{sec:conclusion}.

\section{Methods and Simulations}
\label{sec:methods}

Here we briefly outline our numerical methods and initial conditions.  We base our results on the simulations described in Paper I and the interested reader is referred there for further details.

\subsection{The Monte Carlo Code}
\label{sec:MC}

We simulate star clusters using a H\'e{}non-type Monte Carlo code \citep{Henon71} with improvements for both global and binary dynamics developed by \cite{Stod82} and \cite{Stod86} as incorporated by \cite{Giersz98}.  In this code star clusters are assumed to be spherically symmetrical with their
global dynamical evolution governed by two-body relaxation.  Using this approximation the orbits of individual stars can be described as plane rosettes defined by their energy, $E$, and angular momentum vector, $\vec{J}$.  The dynamical evolution of the cluster over a time $\Delta t$ can then be
calculated using an appropriate scattering angle chosen by Monte Carlo sampling from the theory of two-body relaxation.  This method has the advantage of requiring only a fixed number of operations per particle and thus scaling as $\mathcal{O}(N^{1-2})$, where $N$ is the number of particles in the simulation, as opposed to the $\mathcal{O}(N^{3-4})$ of direct $N$-body simulations.  This allows us to run simulations with $10^{5}-10^{6}$ particles,  similar to the number of stars in real globular clusters, while still being able to perform large parameter space studies.

Strong few-body interactions are not part of the Monte Carlo approximation and must be modelled separately by incorporating cross-sections for such events to occur and prescriptions to determine their outcomes.  For three-body interactions the probability of a close encounter is calculated according to the
cross-sections found in \cite{Giersz01} and the outcomes according the the prescriptions of \cite{Giersz98}.  The cross-sections for binary-single and
binary-binary interactions are calculated following the method of \cite{GandS03} with the outcome of binary-single interactions determined by the formulae given in \cite{Giersz98}.  The outcomes of binary-binary interactions are based on the numerical few-body scattering experiments of \cite{Mikkola84} as implemented by \cite{Stod86}.  The probabilities for exchanges during binary-single and binary-binary interactions are taken from \cite{HHandM96}.

The code includes treatments for single and binary stellar evolution using the SSE \citep{HPandT00} and BSE \citep{HTandP02} packages \citep{GHandH08}.
These packages include prescriptions for stellar and binary evolution from the zero-age main sequence to the stellar remnant at a variety of metallicities
and include analytic treatments of mass-transfer and perturbed evolution in binary systems.  For our purposes, the primary effect of metallicity is higher-mass BHs at low metallicity due to less efficient line driving of stellar winds on the giant and asymptotic giant branch \citep{BKandB02}.  A study of the resulting BH mass distributions is given in \cite{Belczynski06}.  Prescriptions for natal kicks in supernovae remnants \citep{LandL94} are also included.  They are drawn from a Maxwellian distribution with a peak centred at $\sim 190$ km s$^{-1}$ based on the proper motion studies of \cite{HandP97} and then reduced in proportion to the mass of material accreted during fallback as calculated in \cite{BKandB02}.  An estimate for the gravitational wave inspiral timescale based on the quadrupole approximation \citep{Peters64} is also included.  Tidal truncation of the cluster is treated using the prescriptions of \cite{Baumgardt01} \citep{Giersz01,GHandH08}.

The code agrees well with direct $N$-body simulations in the case of equal-masses \citep{Giersz98}, a mass function \citep{Giersz01,Giersz06} and when stellar evolution has been included \citep{GHandH08}.  Simulations using this code have also been shown to re-produce the physical parameters of the observed star clusters M67 \citep{GHandH08}, M4 \citep{GandH08} and NGC 6397 \citep{GandH09}.  Despite these successes the code has some limitations.  Of particular interest for our study is the Spitzer instability \citep{Spitzer87} where massive objects, such as BHs, can fall out of energy equipartition and form a strongly interacting subsystem in the cluster core.  In such a situation the assumption of weak scattering may no longer hold and the Monte Carlo approximation can break down.  \cite{HandG09} have compared the behaviour of the core region of Monte Carlo and direct $N$-body simulations of NGC 6397 and have found good agreement between both sets of simulations in the binary binding energies and escape rates.  This indicates that core dynamics are being re-produced properly.  Artificially small bound subsystems can also be detected by sudden drops in the central potential of the cluster.  None of our simulations demonstrate such an effect, indicating that there is no major unphysical behaviour in the cluster cores.  Finally we showed that the binding energy of the escapers was in good agreement with that found in \cite{OLeary06} (and \cite{SPZandM00}) indicating that the energetics of binaries in the core are being treated accurately.  A more serious concern is the lack of direct few-body integration in our binary-single and binary-binary interactions.  This reduces the number of possible outcomes for these interactions and eliminates both the possibility of mergers during the interaction and the formation of hierarchical triples that can pump up the  eccentricity of the inner binary through resonances \citep{Kozai62} and enhance gravitational wave emission.  Eccentricities are also not generated self-consistently in interactions but are choose randomly.  Since most interactions will tend to favour the formation of high eccentricity orbits, this means that our eccentricities will be under-estimated.  The combination of these effects means that our merger rates will be lower limits and could increase with a more accurate treatment of few-body interactions.

\begin{figure*}
\centering
\includegraphics[width=\textwidth]{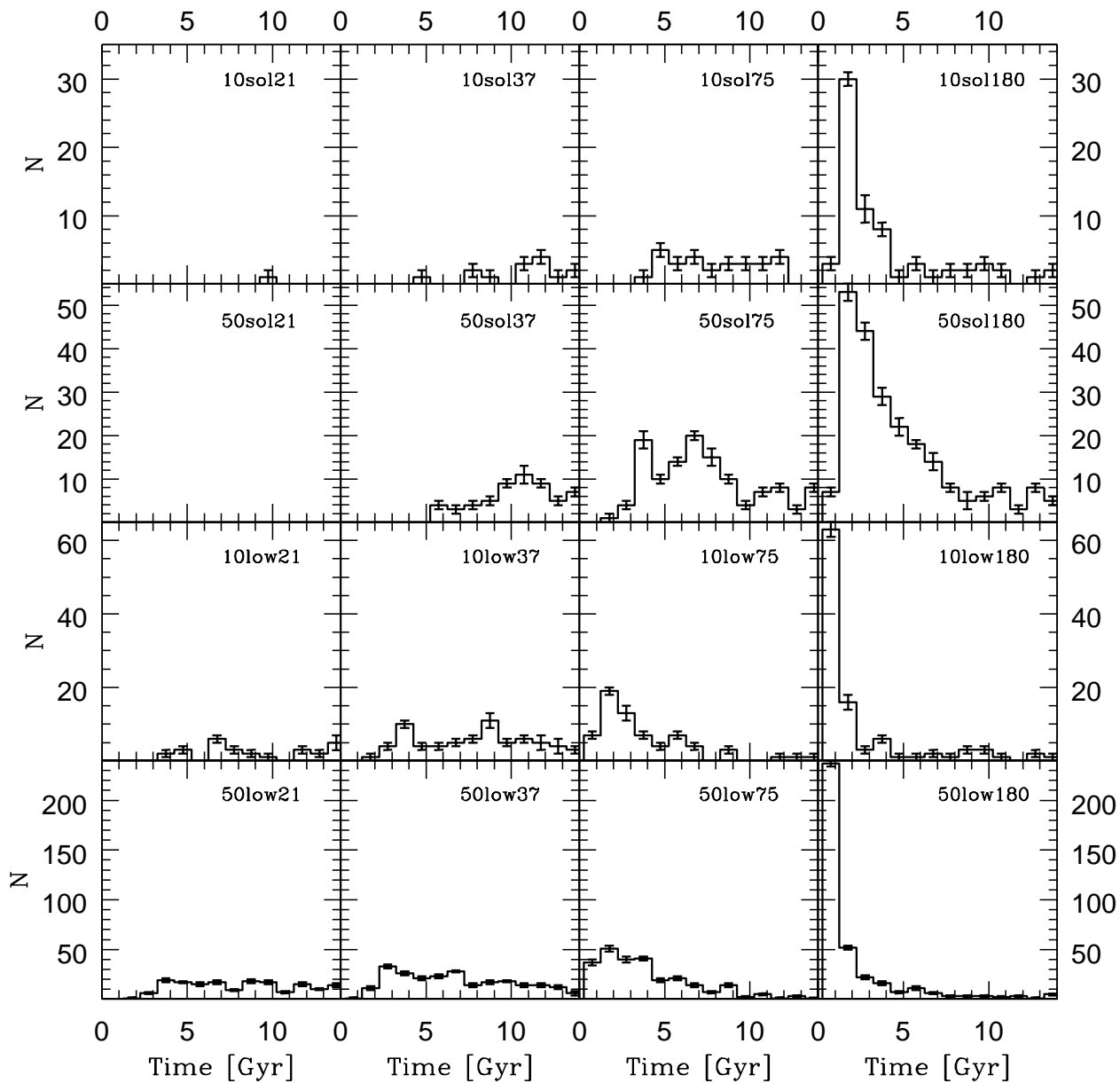}
\caption[BH-BH Escape Rates]{The number of BH-BH escapers per Gyr summed over all ten independent realisations of each set of initial conditions.  The   error bars gives the rms scatter between the independent realisations.\label{fig:EscRate}}
\end{figure*}
\begin{table}
\centering
\caption[Number of BH-BH Escapers and Mergers]{The number of BH-BH binaries that escape from the cluster and the number of BH-BH mergers.  The first column identifies the initial conditions.  The second column gives the total number of BH-BH binary escapers summed over all ten independent realisations of each simulation.   The third column gives the number of BH-BH escapers averaged over all ten independent realisations.  The fourth column gives the number of BH-BH escapers that merge within a Hubble time summed over all ten independent realisations.  The fifth column gives the number of BH-BH escapers that merge averaged over all ten independent realisations.  The uncertainty in columns three and five is the rms scatter across the ten independent realisations.\label{tab:EscMerge}}
\scriptsize{
\begin{tabular}[c]{l r r r r}
\hline
\multicolumn{5}{c}{BH-BH Escapers}\\
\hline
Simulation & $N_{E}$ & $\langle N_{E} \rangle \pm \sigma_{N_{E}}$ & $N_{M}$ & $\langle N_{M} \rangle \pm \sigma_{N_{M}}$ \\
\hline
10sol21  &   1 &  $0 \pm 1$ &   0 &  $0 \pm 0$ \\
10sol37  &  17 &  $2 \pm 1$ &   1 &  $0 \pm 1$ \\
10sol75  &  33 &  $3 \pm 1$ &   5 &  $1 \pm 1$ \\
10sol180 &  69 &  $7 \pm 2$ &  24 &  $2 \pm 1$ \\
50sol21  &   0 &  $0 \pm 0$ &   0 &  $0 \pm 0$ \\
50sol37  &  61 &  $6 \pm 2$ &   2 &  $0 \pm 1$ \\
50sol75  & 131 & $13 \pm 2$ &  16 &  $2 \pm 1$ \\
50sol180 & 233 & $23 \pm 3$ &  36 &  $4 \pm 1$ \\
10low21  &  32 &  $3 \pm 1$ &   3 &  $0 \pm 1$ \\
10low37  &  72 &  $7 \pm 1$ &   8 &  $1 \pm 1$ \\
10low75  &  68 &  $7 \pm 1$ &  25 &  $3 \pm 2$ \\
10low180 & 104 & $10 \pm 2$ &  58 &  $6 \pm 1$ \\
50low21  & 165 & $17 \pm 2$ &   2 &  $0 \pm 1$ \\
50low37  & 243 & $24 \pm 5$ &  26 &  $3 \pm 2$ \\
50low75  & 260 & $26 \pm 4$ &  54 &  $5 \pm 2$ \\
50low180 & 372 & $37 \pm 2$ & 161 & $16 \pm 3$ \\
\hline
\end{tabular}
}
\end{table}

\subsection{Initial Conditions}
\label{sec:InitCond}

The initial conditions for our simulations were described in Paper I and we merely summarise them here.  Our results are based on a set of 160 simulations,
each with $N = 5 \times 10^{5}$ particles, a Kroupa power-law IMF \citep{KroupaIMF} with a lower slope of $\alpha_{l} = 1.3$, an upper slope of $\alpha_{l} = 2.3$, a break mass of 0.5 M$_{\odot}$ and masses between 0.1 M$_{\odot}$ and 150 M$_{\odot}$.  The models are initialised with Plummer profiles with a tidal radius of $r_{t} = 150$ pc.  We have used two initial binary fractions ($f_{b}$), $10\%$ and $50\%$, with initial binary parameters taken from \cite{KroupaBin}.  We use two metallicities, $Z = 0.02$ (solar) and $Z = 0.001$ (low).  Finally we use four different initial concentrations defined by the ratio of the tidal radius to the initial half-mass radius ($r_{h}$), $r_{t}/r_{h} \in {21,37,75,180}$, corresponding roughly to initial number densities within the half-mass radius of ${10^{2},10^{3},10^{4},10^{5}}$ pc$^{-3}$.  Smaller half-mass radii reduce the half-mass relaxation time \citep{Spitzer87}, 
\begin{equation}
t_{rh} = 0.138\frac{N^{1/2}r_{h}^{3/2}}{\langle m
  \rangle^{1/2}G^{1/2}\ln{\gamma N}}
\end{equation}
where $\langle m \rangle$ is the average mass in the system, $G$ is Newton's gravitational constant and $\gamma = 0.02$ is an empirically determined constant, in the highest concentration clusters and thus leads to faster dynamical evolution.  Taken together, this gives 16 possible combinations of initial conditions.  We have then performed ten independent realisation of each simulation in order to constrain statistical fluctuations for a total of 160 individual simulations.  We confirmed that ten simulations is sufficient for accurate statistics.  The initial conditions are summarised in Table 1 of Paper I.  The simulations have been performed on the HLRS supercomputer in Stuttgart.  Each is run on a single processor and it takes between 4 and 24 hours to simulate a cluster for 1 $T_{H}$.

\begin{figure}
\centering
\includegraphics[width=0.5\textwidth]{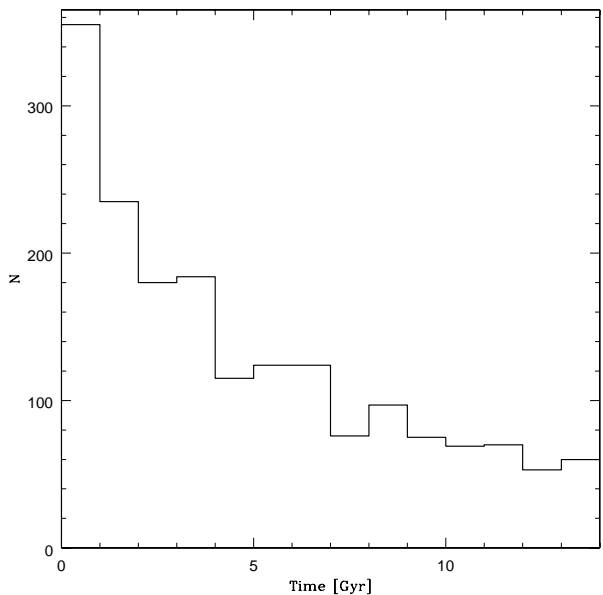}
\caption[Total BH-BH Escape Rate]{The total number of BH-BH escapers per Gyr summed over all 160 simulations.\label{fig:TotEsc}}
\end{figure}

\section{Properties of Escaping Binaries}
\label{sec:properties}

In Paper I we discovered no BH-BH mergers within our clusters although we found a few potential LISA sources.  We found, however, that many of our clusters ejected hard BH-BH binaries with binding energies  $> 1000k_{B}T$ where $k_{B}T = M_{\rm core}\sigma_{\rm core}^{2}/2N_{\rm core}$ is the
thermal energy in the core of the cluster, $M_{\rm core}$ is the total mass of the core, $N_{\rm core}$ is the number of stars in the core and $\sigma_{\rm core}$ is the velocity dispersion in the core.  This result is consistent with the binding energy distribution found by both \cite{SPZandM00} and \cite{OLeary06}.  We speculated that these hard binaries may merge within a Hubble time and be sources of gravitational waves for both ground- and space-based detectors.

\begin{figure*}
\centering
\includegraphics[width=\textwidth]{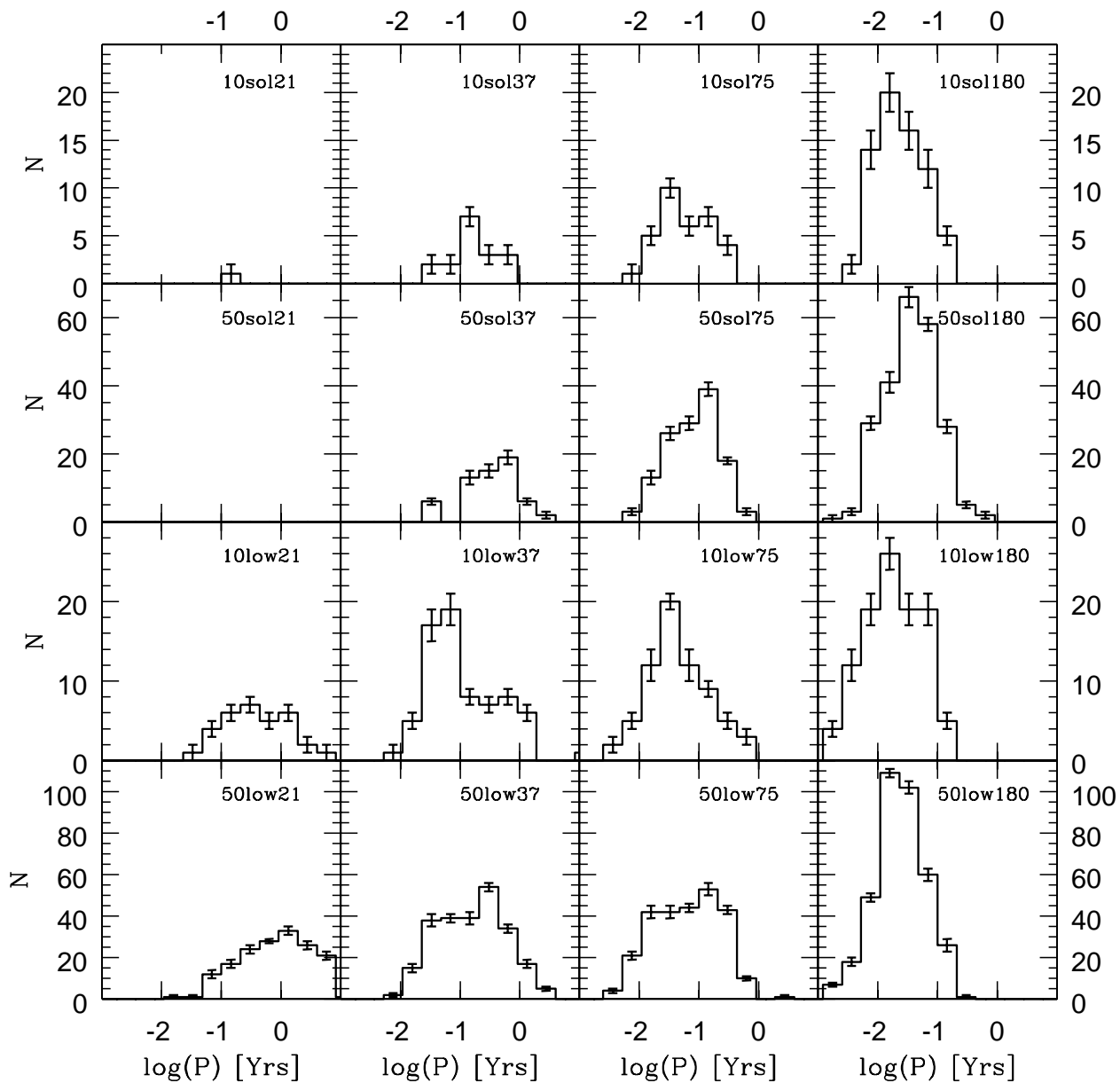}
\caption[Period Distribution of BH-BH Escapers]{The period distribution of BH-BH escapers binned uniformly in log space and summed over all ten  independent realisations of each set of initial conditions.  Error bars give the rms scatter between the ten realisations.\label{fig:Periods}}
\end{figure*}

In Table~\ref{tab:EscMerge} we present the number of escapers from each of our simulation both summed and averaged over all ten independent realisations.  The trends in Table~\ref{tab:EscMerge} follow similar trends to those found for the formation of BH-BH binaries in Table 3 of Paper I, namely more BH-BH
escapers at high $f_{b}$, more BH-BH escapers at higher initial concentration and more BH-BH escapers at low metallicity.  The trend with $f_{b}$ is due 
both to the larger number of stars and thus the larger number of BHs in high-$f_{b}$ simulations and to the larger number of binaries available for BHs to be exchanged into.  The trend with initial concentration is also clear; at higher density there are more interactions per unit time and hence more
dynamical BH-BH binary formation and ejection over 1 $T_{H}$.  The trend with metallicity is due both to the larger number of BHs and their greater mass at low-$Z$.  The mass-segregation timescale governing how quickly massive objects sink to the cluster centre is given by:
\begin{equation}
t_{ms} \propto t_{rh} \frac{m_{2}}{m_{1}}
\end{equation}
where $m_{1} > m_{2}$ \citep{WJandR00,KAandS07} and the more massive BHs and BH-BH binaries in low-metallicity clusters mass-segregate faster.  Thus the
massive BHs enter the high density cluster cores more quickly than their lower-mass counterparts and the dynamical evolution is accelerated.  Column
two of Table~\ref{tab:EscMerge} shows that that there is very little simulation-to-simulation scatter in the number of BH-BH binaries ejected.

\begin{figure*}
\centering
\includegraphics[width=\textwidth]{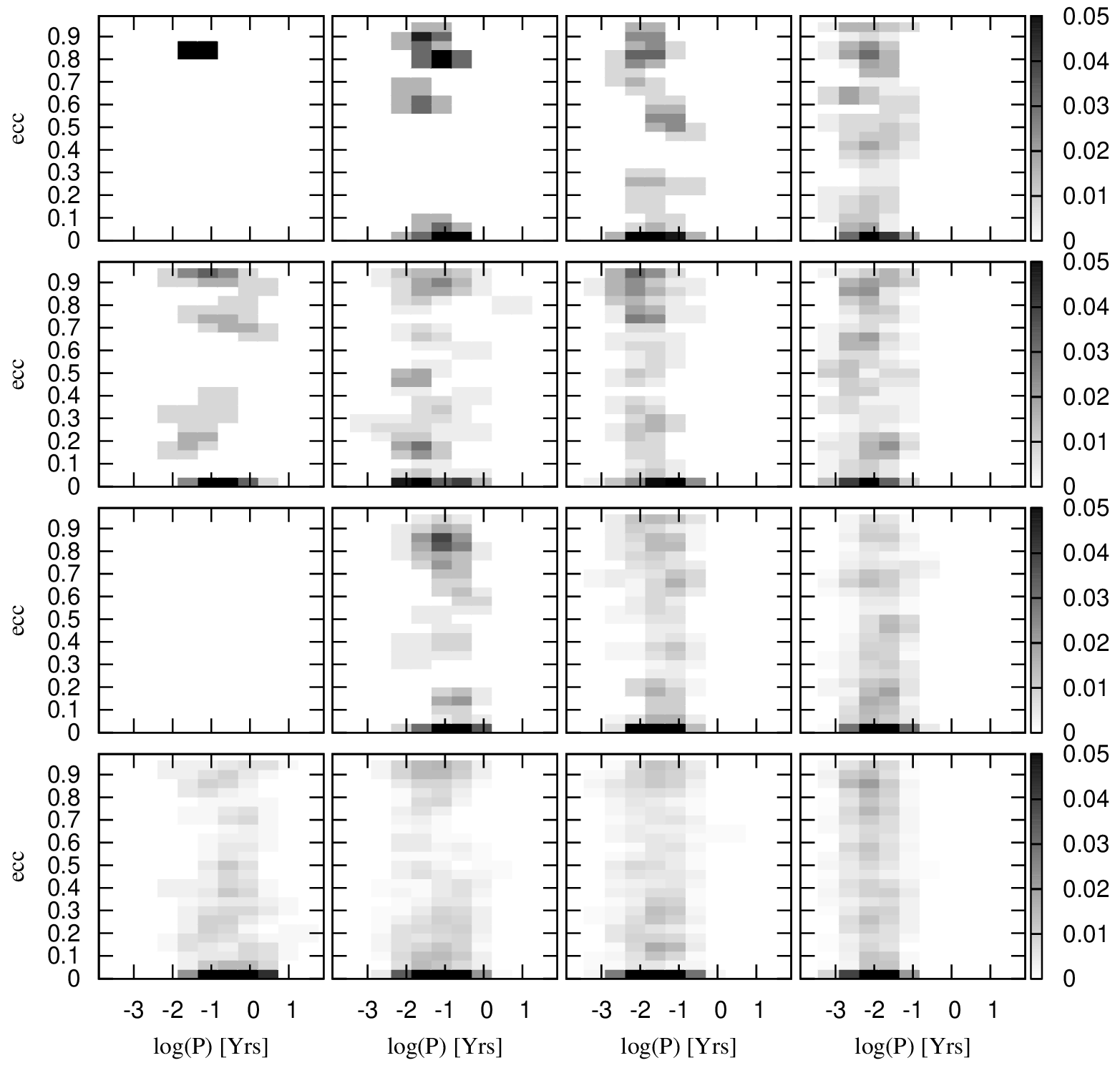}
\caption[Period vs. Eccentricity for BH-BH Escapers]{The eccentricity as a function of period for BH-BH binaries at the time of escape from the cluster.  Each panel shows all binaries from all all independent realisations of a single set of initial conditions.  From top to bottom: $f_{b} = 0.1$ and $Z = 0.02$, $f_{b} = 0.5$ and $Z = 0.02$, $f_{b} = 0.1$ and $Z = 0.001$, $f_{b} = 0.5$ and $Z= 0.001$.  From left to right $r_{t}/r_{h} =$ 21, 37, 75, 180.\label{fig:PvsEcc}}
\end{figure*}

Figure~\ref{fig:EscRate}, which gives the number of BH-BH escapers per Gyr relative to the start of the simulation from which they escaped, confirms this
picture.  There are more BH-BH escapers at high $f_{b}$, high initial concentration and low metallicity.  Figure~\ref{fig:EscRate} also shows the effect the initial conditions have on the time and rate at which BH-BH binaries escape from the simulations.  For low initial concentrations and solar metallicity the escape of BH-BH binaries commences late in the evolution of the cluster and happens at a slow, fairly constant rate.  By contrast at high initial concentrations and low metallicity there is a large burst of escapers very early but the escape rate drops-off later in the life of the cluster.  This is due to the shorter relaxation time and mass-segregation timescale in these simulations.  In the high concentration, low metallicity simulations the BHs and binaries mass segregate and sink to the high density core more swiftly where they interact and are ejected rapidly.  This process proceeds more slowly in the low initial concentration, high metallicity clusters and therefore these clusters produce fewer ejections within a Hubble time.  The initial binary fraction does not affect the location of peak BH-BH ejection but does affect the overall number of BH-BH binaries ejected and clusters with a higher initial binary fraction eject more BH-BH binaries than their lower $f_{b}$ counterparts.  In figure~\ref{fig:TotEsc} we present the total number of  escapers summed over all 160 simulations and binned by Gyr assuming all clusters form at the same time.  It is apparent that the escape rate will be
highest early in the life of the cluster population where it is dominated by low metallicity, high density clusters but that there will be an appreciable
number of escapers at late time as well.

\begin{figure*}
\centering
\includegraphics[width=\textwidth]{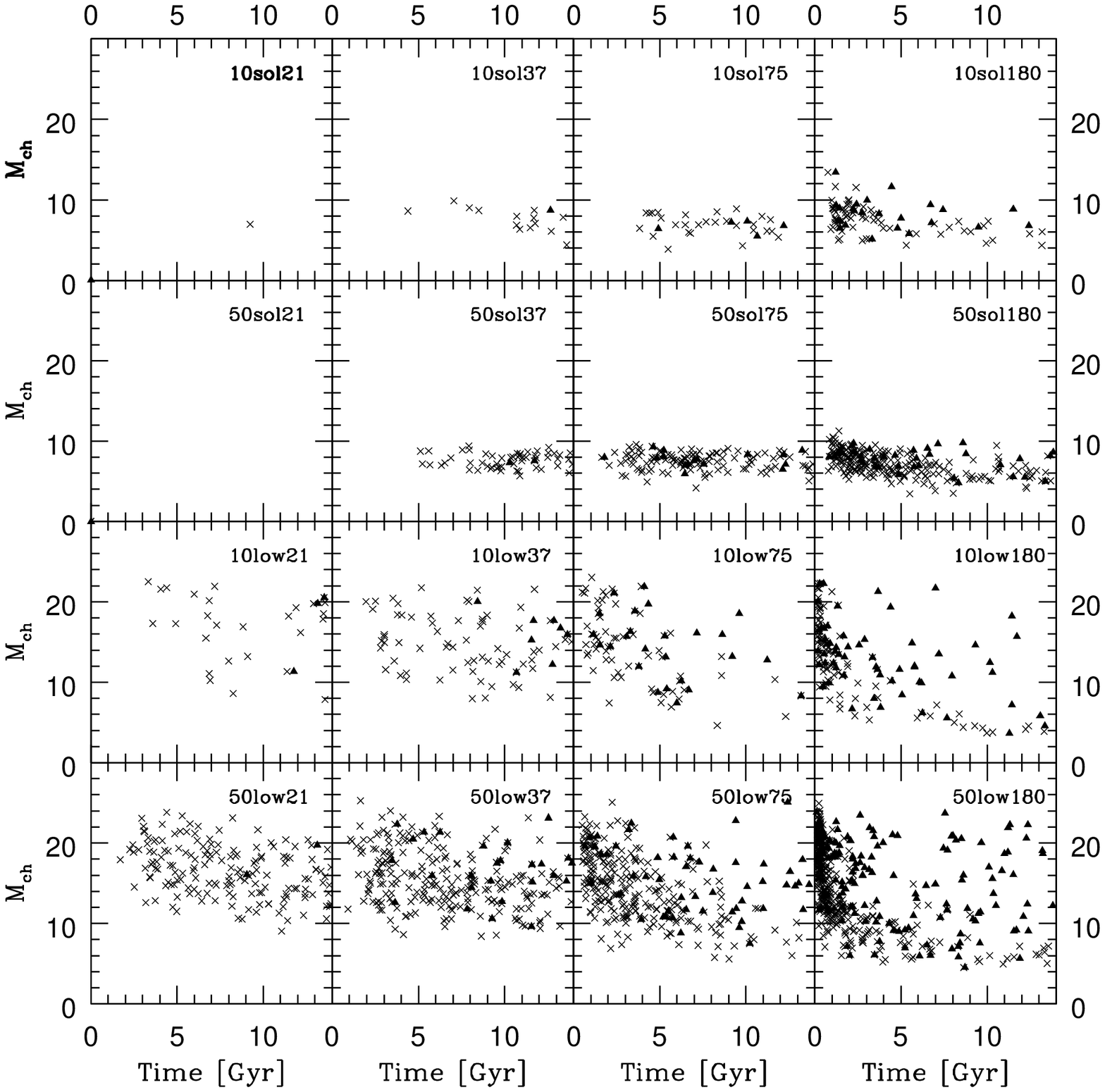}
\caption[Chirp Masses of Escapers and Mergers]{$M_{\rm chirp}$ for BH-BH binaries at the time of escape from the cluster (crosses) and BH-BH mergers   (solid triangles).  Each panel shows all binaries in all ten independent realisations for a single set of initial conditions.\label{fig:Mchirp}}
\end{figure*}

Figure~\ref{fig:Periods} gives the period distribution of all escaping binaries.  All periods are short with few more than a year and many less than a day.  Binaries with periods of less than a few days are capable of merging within a Hubble time due to gravitational radiation
\citep{Peters64} and are thus potential sources for ground-based detectors.  Furthermore, unlike the binaries that remain within the cluster, these binaries
will not be subject to disruption in the further interactions that prevent mergers within the cluster as described in Paper I.  There is little variation
in the period distribution with cluster parameters other than a tendency towards shorter periods in high density clusters.  This is because the core velocity dispersion and thus the core kinetic energy is higher in these clusters meaning that a binary must be more energetic and thus shorter period in these clusters in order to survive and be ejected.

\begin{figure*}
\centering
\includegraphics[width=\textwidth]{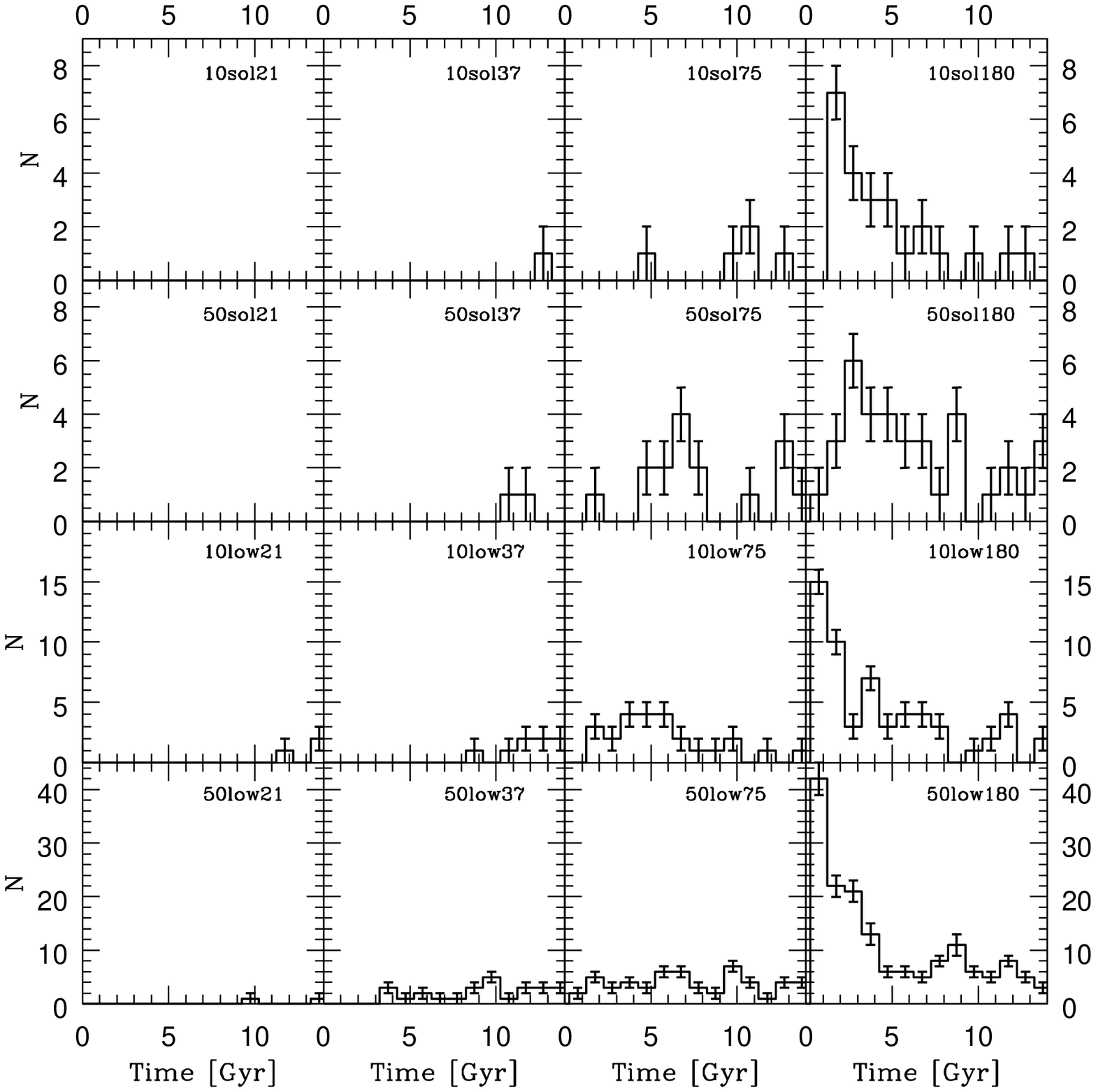}
\caption[Individual BH-BH Merger Rates]{Merger rates for BH-BH escapers binned per Gyr.  Each panel gives the merger rate for one choice of initial   conditions summed over all ten independent realisations.  The error bars give the rms scatter between the independent realisations.\label{fig:MergeRate}}
\end{figure*}
\begin{figure}
\centering
\includegraphics[width=0.5\textwidth]{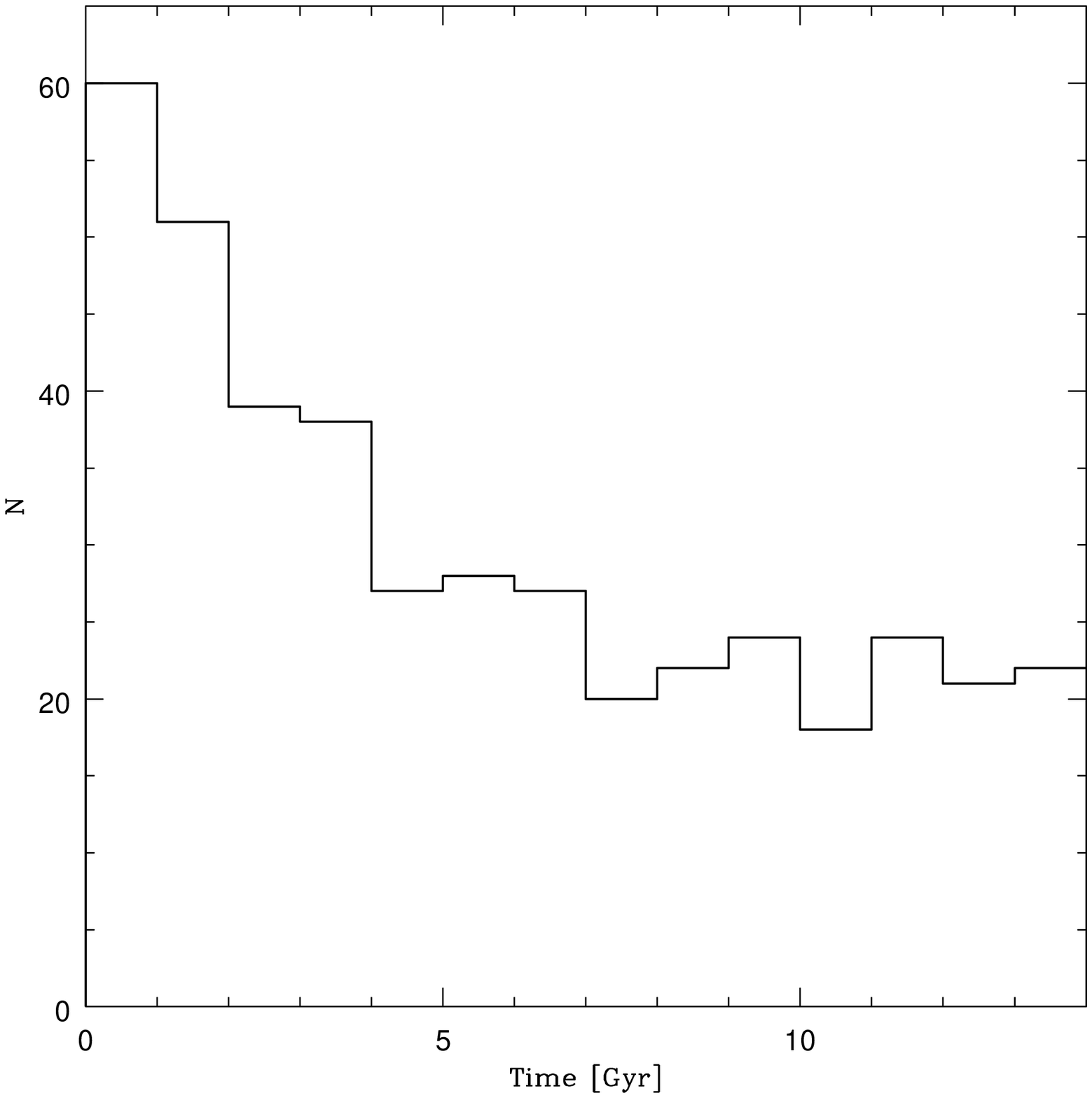}
\caption[Total BH-BH Merger Rate]{The number of BH-BH merger binned per Gyr and summed over 160 all realisations of all initial conditions.\label{fig:TotMerge}}
\end{figure}

In figure~\ref{fig:PvsEcc} we present the eccentricity of escaping BH-BH binaries as a function of their period at the time of escape from the cluster.
High eccentricity can substantially enhance the power radiated in gravitational waves \citep{PandM63,Peters64} and thus high eccentricity binaries will have shorter merger times than their circular counterparts.  Eccentric binaries also emit gravitational waves in higher harmonics of the orbital frequency \citep{PandM63} and this can move long-period binaries into the LISA band.  Figure~\ref{fig:PvsEcc} shows that our simulations produce BH-BH escapers with a large range of eccentricities at all periods indicating a rather high probability of short-period, eccentric binaries.  These eccentricities must be treated with some caution since they are not self-consistently generated but rather the product of a random choice at the end of an interaction.  In  reality, however, interactions tend to produce high eccentricity binaries so if anything figure~\ref{fig:PvsEcc} underestimates the average eccentricity to be expected in the BH-BH escapers.  As we will demonstrate in \S~\ref{sec:mergers} many of the binaries have a period and eccentricity combination that will lead to a merger within a Hubble time and thus we predict that star clusters will enrich galactic fields with BH-BH merger candidates.  We also note that some of these binaries may be in the LISA band with significant eccentricity upon ejection and thus we predict that escaping BH-BH binaries from star clusters may be eccentric LISA sources.  We will investigate this possibility further in \S~\ref{sec:LISAdet}.

Finally in figure~\ref{fig:Mchirp} we show the chirp masses $M_{\rm chirp}$ of all BH-BH binaries at the time of escape as well as for all BH-BH mergers (see \S~\ref{sec:mergers}).  The chirp mass,
\begin{equation}
M_{\rm chirp} = \frac{(m_{1}m_{2})^{3/5}}{(m_{1}+m_{2})^{1/5}}
\end{equation}
where $m_{1}$ is the mass of the primary and $m_{2}$ is the mass of the secondary, is an important quantity in gravitational wave studies since the amplitude of a gravitational wave, $h_{0}$, is proportional to $M_{\rm chirp}$
\citep{Pierro01}:
\begin{equation}
h_{0} \propto \frac{G^{5/3}\omega^{2/3}M_{\rm chirp}^{5/3}}{rc^{4}}
\end{equation}
where $\omega$ is the angular frequency of the binary, $r$ is the distance from the binary to the observer, and $c$ is the speed of light.  Thus it is 
this quantity rather than the total mass that is important for gravitational wave detection.  $M_{\rm chirp}$ is larger for the low metallicity systems due
to the afore mentioned higher mass of BHs derived from lower metallicity progenitors.  It is also clear that in the low metallicity cases the binaries with the highest chirp mass escape first.  This is another consequence of mass segregation since the high mass BH-BH binaries in the core will interact and be ejected more rapidly than their lower-mass counterparts.  This is not apparent in the high metallicity cases since there the spread in mass is not sufficient for this effect to be important (see Paper I).  It is worth noting, however, that the time-dependence of escaper chirp mass does not translate
into a time-dependent merger chirp mass.  This is due to the time lag between the escape of a BH-BH binary and its merger in the field since a short period,
low mass binary may still merge before a long-period, high mass binary.  This will be explored in \S~\ref{sec:mergers}.

\section{BH-BH Mergers}
\label{sec:mergers}

Here we calculate the number of escaping BH-BH binaries that will merge within a Hubble time and when the merger will occur.  We use the formalism presented in \cite{PandM63} and \cite{Peters64} in order to calculate the gravitational wave inspiral timescale for BH-BH binaries in a galactic field.  In this approximation the field equations are linearised and the gravitational wave radiation is then calculated in a multipole expansion.  The first non-zero order in this expansion is the quadrupole term, which can be used to calculate the energy and angular momentum carried away from the binary by gravitational waves.  This can then be used to calculate the effect of gravitational radiation on the orbital elements of the binary.  The orbit-averaged change in semi-major axis, $a$, and eccentricity, $e$, in the quadrupole approximation are:
\begin{equation}
  \label{eq:PandMa}
  \langle \dot{a} \rangle = -\frac{64}{5}\frac{G^{3}}{c^{5}}
  \frac{m_{1}m_{2}(m_{1}+m_{2})}{a^{3}(1-e^{2})^{7/2}} \left( 1 +
  \frac{73}{24}e^{2} + \frac{37}{96}e^{4} \right)
\end{equation}
and
\begin{equation}
  \label{eq:PandMe}
  \langle \dot{e} \rangle = -\frac{304}{15}e\frac{G^{3}}{c^{5}}
  \frac{m_{1}m_{2}(m_{1}+m_{2})}{a^{4}(1-e^{2})^{5/2}} \left( 1 +
  \frac{121}{304}e^{2} \right)
\end{equation}
\citep{PandM63,Peters64}.  For each escaping binary we solve equations~\ref{eq:PandMa} and~\ref{eq:PandMe} using a 4$^{\rm th}$-order Runge-Kutta integrator with time steps chosen such that there is never more than a $1\%$ change in $a$.  We integrate from the time of escape from the cluster until either the binaries merge or 1 $T_{H}$ is reached.  This gives us both an inspiral timescale for the binary and, after each timestep, a self-consistently generated, quasi-static set of orbital parameters that can be used to calculate a gravitational wave signal for LISA using the formalism of \cite{Pierro01} as will be discussed in more detail in \S~\ref{sec:LISAdet}.

In table~\ref{tab:EscMerge} we show the number of escaping BH-BH binaries that merge within a Hubble time.  The total number of BH-BH mergers tracks the
total number of BH-BH escapers however the proportion of escaping BH-BH binaries that merge within a Hubble time increases in clusters with low
metallicity and high initial concentration.  The correlation with metallicity is again due to the larger mass of BHs in the low metallicity simulations and
the correspondingly stronger gravitational wave radiation.  The correlation with initial concentration is due to the fact that binaries tend to have
shorter periods in clusters with a high concentration and thus have a shorter distance over which to inspiral and merge.  Again there is relatively little
simulation-to-simulation scatter in the number of mergers.

Figure~\ref{fig:MergeRate} shows the number of mergers for each set of initial conditions per Gyr where the time of merger of a BH-BH binary is given relative to the start of the simulation that produced it.  In this plot we present a sum rather than an average over each set of initial conditions.  This is partly due to the small number of mergers but also because once the BH-BH binaries escape from their parent clusters they will form a single population in a galactic field.  The overall shapes broadly follow those found for the number of escapers per Gyr in figure~\ref{fig:EscRate} but with a time delay between when escapers and mergers begin and when the peak number of escapers and  mergers per Gyr is found.  The mergers are delayed because the inspiral in a galactic field is not an instantaneous process and it often takes a binary several Gyrs to merge.  Again assuming that all star clusters form at the same time early in the life of the universe then the merger rate in the early universe will be dominated by BH-BH binaries originating from dense, low-metallicity clusters whereas at the current age of the universe clusters of all types will be contributing mergers in similar numbers.  In figure~\ref{fig:TotMerge} we present the total number of mergers per Gyr summed over all realisations of all initial conditions.  The profile is flatter than for the total number of escapers in figure~\ref{fig:TotEsc} but still shows a peak in the merger rate while the cluster population is young.  This implies that if we can detect BH-BH mergers at large redshift then mergers from young clusters in the early universe will dominate the detection rate.  It also indicates that if dense, low-metallicity clusters form later in the universe they will be major contributors to the local BH-BH merger rate.

\begin{figure*}
\centering
\includegraphics[width=\textwidth]{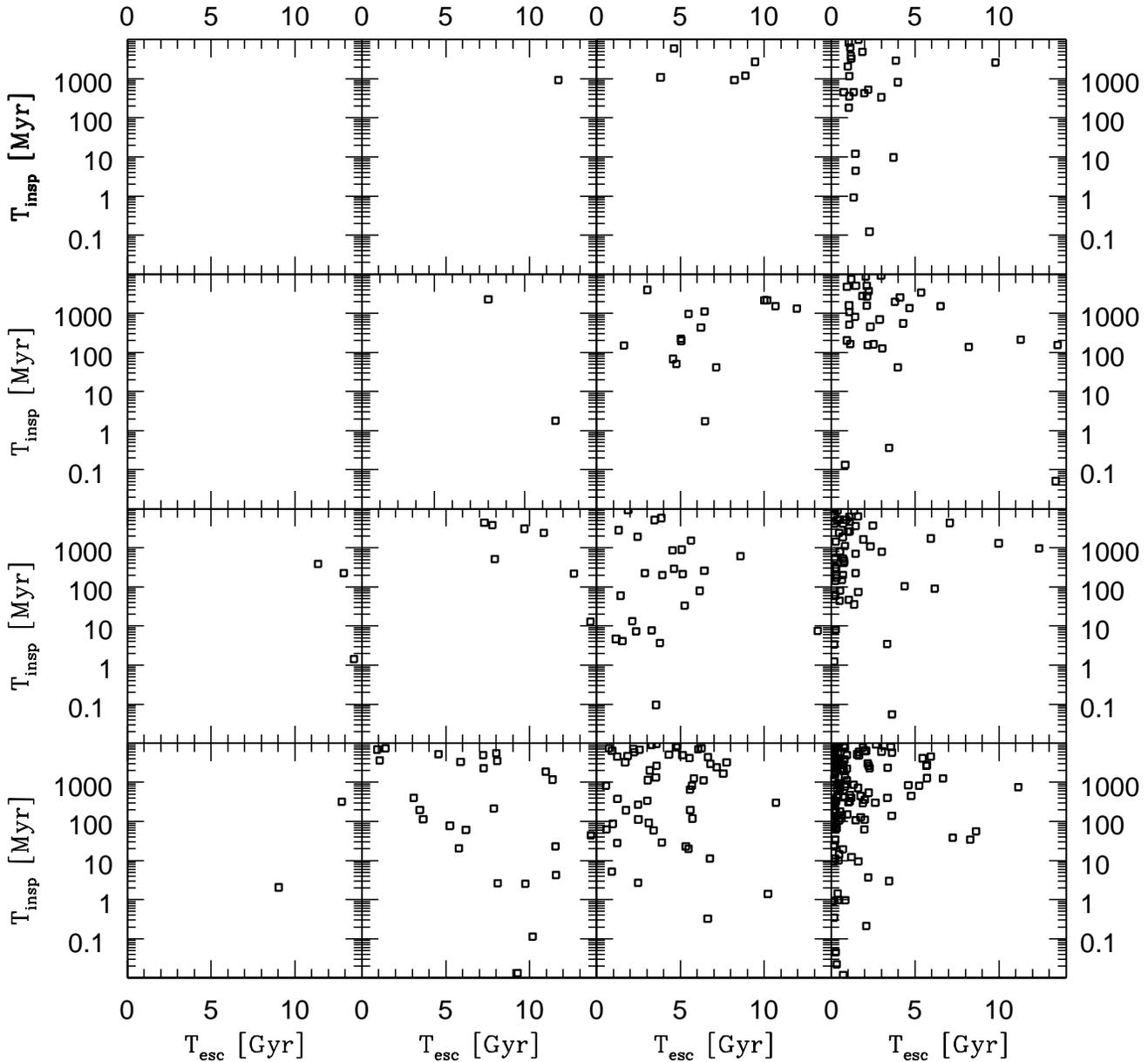}
\caption[Merger Timescales]{The merger timescale as a function of escape time for all escaping BH-BH binaries that merge.  Each panel gives merger   timescales for all ten independent realisations of each set of initial conditions.\label{fig:Tinsp}}
\end{figure*}

Figure~\ref{fig:Tinsp} give the inspiral timescale for each BH-BH binary from the time of escape from the cluster until the merger as calculated by equations~\ref{eq:PandMa} and~\ref{eq:PandMe} and as a function of time of escape from the cluster.  There are no particular trends in inspiral time with
escape time in any of the simulations.  This implies that there is no clear link between merger time and escape time and that it is probably impossible to
derive an escape rate from an observed merger rate.  It also explains why there is no evolution in $M_{\rm chirp}$ of mergers as a
function of time.  High $M_{\rm chirp}$ binaries with a relatively long period that escape early are as likely as low $M_{\rm chirp}$ binaries with a short
period that escape late.  Thus the trend in the early escape of high $M_{\rm chirp}$ binaries does not translate into a similar trend in the merger rate.

\begin{figure*}
\centering
\includegraphics[width=\textwidth]{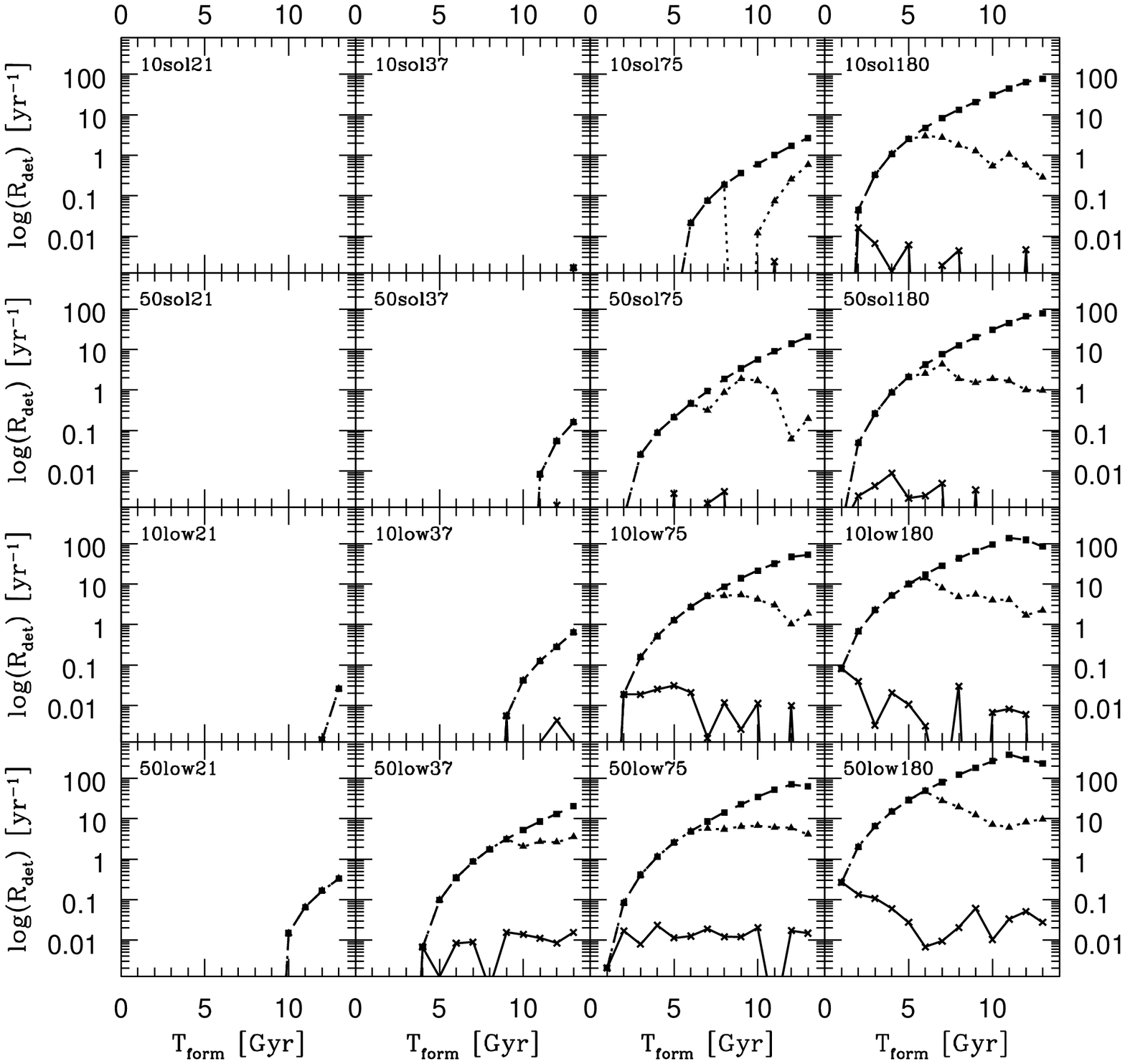}
\caption[Detection Rates for by Cluster Type]{Detection rates by cluster type.  Each panel gives the expected detection rate if the entire cluster  population in the universe was composed of identical clusters, each with the corresponding initial conditions.  The x-axis gives the lookback time to $T_{\rm form}$ in  Gyr.  The solid line with crosses is for $D_{L,0} = 19.1$ Mpc, the dotted line with triangles is for $D_{L,0} = 191.0$ Mpc and the dashed line with   squares is for $D_{L,0} = 1910.0$ Mpc.\label{fig:RatesSingle}}
\end{figure*}

\section{Detection Rates for Ground-Based Detectors}
\label{sec:LIGOdet}

Now we must determine if the BH-BH mergers that are produced by our simulations can be detected by gravitational wave observatories.  We follow the approach of \cite{OLeary06} who estimate the detectability of a BH-BH merger by comparing it to a template merger with a known signal-to-noise ratio
at a given distance.  A merger with a red-shifted chirp mass of $\mathcal{M}_{\rm chirp} = (1+z_{m})M_{\rm chirp}$, where $z_{m}$ is the redshift of the merger, is considered detectable at a luminosity distance $D_{L} = (1+z_{m})D_{\rm prop}$, where $D_{\rm prop}$ is the proper distance to the merger, if \citep{OLeary06}:
\begin{equation}
\label{eq:detectability}
\frac{D_{L,0}}{D_{L}} \left( \frac{\mathcal{M}_{\rm chirp}}{\mathcal{M}_{\rm chirp,0}} \right)^{5/6} \sqrt{ \frac{s(f_{\rm off})}{s(f_{\rm off,0})} } > 1
\end{equation}
where $D_{L,0}$ is the luminosity distance at which a merger of red-shifted chirp mass $\mathcal{M}_{\rm chirp,0}$ is known to be detectable at a given
signal to noise ratio.  For the planned advanced LIGO detector it is estimated that a equal-mass, NS-NS binary with $M_{\rm chirp} = 1.2 {\rm M}_{\odot}$ can be detected with a signal-to-noise ratio of 8 at a distance of 191 Mpc \citep{OShaughnessy05,Harry05}.  The distance for the current generation of detectors is some ten times lower.  The detector response function, $s(f_{\rm off})$ is given by \citep{OLeary06}:
\begin{equation}
\label{eq:response}
s(f_{\rm off}) = \int_{0}^{f_{\rm off}} \frac{(f^{\prime})^{-7/3}}{S_{N}(f^{\prime})} df^{\prime}
\end{equation}
where $S_{N}(f^{\prime})$ is the noise spectrum of the detector and is approximated as \citep{CandF94,OLeary06}:
\begin{equation}
\label{eq:noise}
S_{N}(f^{\prime}) \propto \left\{ \begin{array}{ll} \infty , & f^{\prime} < 10 {\rm Hz} \\ \left( \frac{f_{0}}{f^{\prime}} \right)^{4} + 2 \left[ 1 +   \left( \frac{f^{\prime}}{f_{0}} \right)^{2} \right] , & f^{\prime} \ge 10 {\rm Hz} \end{array} \right.
\end{equation}
with $f_{0} = 70$ Hz.  A constant of proportionality is not required in equation~\ref{eq:noise} because it will cancel out in equation~\ref{eq:detectability}.

The cut-off frequency, $f_{\rm off}$ is the frequency at which the merger occurs and detailed relativistic modelling becomes necessary.  For the reference inspiral we use the estimate of \cite{CandF94} which yields $f_{\rm off,0} \sim 720$ Hz for the NS-NS reference binary we have chosen.  For the BH-BH inspiral the frequency is generally taken to be that of the last circular orbit before the final plunge.  We use the same circular estimate as \cite{OLeary06}:
\begin{equation}
\label{eq:foff}
f_{\rm off} \approx 200 \left( \frac{20 \rm{M}_{\odot}}{M} \right) \left( \frac{1}{1+z_{m}} \right) {\rm Hz}
\end{equation}
where $M$ is the total mass of the inspiralling binary.

From our simulations we have $M_{\rm chirp}$ for each of our mergers and the time of the merger relative to the start of the simulation.  By assuming a formation time, $T_{\rm form}$, for each simulation relative to the age of the universe we can calculate a lookback time, redshift and proper distance to the merger and use equation~\ref{eq:detectability} to determine if the merger can be detected.  To convert this to a merger rate per year for ground-based detectors we must make assumptions about the formation history and density of globular clusters in the universe.  For simplicity we assume
that all globular clusters form in a single burst at some time $T_{\rm form}$ in the past.  Again for simplicity we assume a constant number density of
$n_{0} \approx 8.4 h^{3}$ clusters Mpc$^{-3}$.  This estimate comes from \cite{SPZandM00} and is based on observations of the local universe taking into account the different specific frequencies of clusters in galaxies of different Hubble types.  We then bin the detections in uniform bins in lookback time, calculate the proper distance, $D_{{\rm prop},i}$, to the edge of each bin, and use the equation \citep{OLeary06}
\begin{equation}
\label{eq:rate}
R_{\rm det} = \sum_{i = 1}^{100} \frac{N_{i}}{\Delta t} \frac{4\pi}{3} n_{0} (D_{{\rm prop},i}^{3} - D_{{\rm prop},i-1}^{3}) (1+z_{i})^{-1}
\end{equation}
to calculate the detection rate.  $N_{i}$ is the number of detections calculated from our simulation in bin $i$, $\Delta t$ is the bin width and the factor $(1+z_{i})^{-1}$ comes from the cosmological time dilation of the detection rate.  We use the cosmological parameters $h = 0.72$, $\Omega_{m} = 0.27$ and $\Omega_{Lambda} = 0.73$.  We note that it would be possible in principle to perform a more careful estimate of the density of globular clusters in the universe by using press-schechter formalism and making assumptions about the baryon fraction per halo, the fraction of baryons that become globular clusters, the evolution of the globular cluster mass function and globular cluster formation rate.  This would, however, introduce a large number of uncontrolled parameters into our modelling and since our simulations do not cover the whole range of globular cluster masses convolving them with the evolution of the globular cluster mass function seems unjustified.  For these reasons we will limit ourselves to simple assumptions about the globular
cluster population in the universe and save more detailed modelling of the cluster population for when we have a larger set of simulations.

In figure~\ref{fig:RatesSingle} we present the detection rate per year as a function of lookback time to the assumed formation event.  We have chosen a lookback time to $T_{\rm form}$ to be between 1 and 13 Gyr.  We have also calculated rates for three different detector sensitivities (parametrised by $D_{L,0}$).  The dotted line represents advanced LIGO with the solid line representing a detector ten times less sensitive (corresponding roughly to the detectors currently in operation) and the dashed line corresponding to a detector ten times more sensitive.  For the current generation of detectors there is a very low probability of detection since the rate seldom exceeds 0.1 detection per year and is often much lower.  The current detectors are not sensitive enough to detect mergers out to cosmological distances and thus probe a volume that is too small to produce a substantial detection rate.  By contrast the prospects for advanced LIGO are much better with most of the high-density, low metallicity clusters promising detection rates of more than
1 yr$^{-1}$ and some of the most optimistic providing $\sim 50$ detections yr$^{-1}$.  The advanced LIGO detection rate demonstrates some interesting trends with formation time.  The low density, high metallicity simulations only produce detections if the clusters are assumed to form very early in the universe.  This is because these clusters evolve slowly and do not produce BH-BH mergers until late in their lives (see figure~\ref{fig:MergeRate}).  By contrast the higher density, lower metallicity clusters can provide a significant detection rate even if they form quite late in the universe.  Indeed these clusters produce a peak detection rate for a $T_{\rm form}$ of only 5-6 Gyr ago since, according to figure~\ref{fig:MergeRate}, they evolve more quickly and produce the majority of their mergers while still young.  Thus they have the potential to contribute significantly to the merger rate while only a few Gyrs old.  The peak at 5-6 Gyr comes form the balance between detection volume and detectability.  For formation times less than 5-6 Gyr ago advanced LIGO will be able to detect all mergers and thus by choosing an earlier formation time the volume of space in which globular clusters exist and thus the number of globular cluster increases, and thus the detection rate goes up.  Clusters that form more than $\sim 6$ Gyr ago are sufficiently distant that advanced LIGO cannot detect all of the mergers and in particular the early peak merger rate moves out of range for LIGO detection.  Thus the detection rate starts to fall again.  For a ten times greater sensitivity (currently well beyond our technical capabilities) we would be able to detect all mergers at much larger distances and thus the detection rate continues to increase with detection volume up to a $T_{\rm form}$ of $\sim 10$ Gyr ago.

\begin{figure}
\centering
\includegraphics[width=0.5\textwidth]{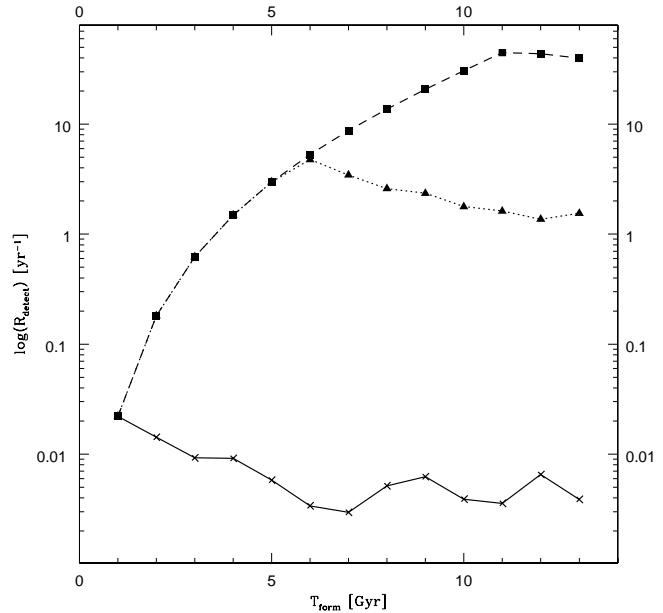}
\caption[Total Detection Rate]{The detection rate assuming all types of clusters are present equally in the universe with an overall number density of $n_{0} = 8.4 h^{3}$ Mpc$^{-3}$.  The solid line with crosses is for $D_{L,0} = 19.1$ Mpc, the dotted line with triangles for $D_{L,0} = 191.0$ Mpc and the dashed line with squares is for $D_{L,0} = 1910.0$ Mpc.\label{fig:TotRate}}.
\end{figure}

Figure~\ref{fig:TotRate} gives the detection rate assuming that all of our simulated clusters are present in the universe in equal numbers.  The shape is very similar to that of the low metallicity, high density clusters, indicating that it is these clusters that will dominate the detection rate.  The overall detection rate is, however, much lower.  The detection rate for the current generation of detectors drops to little more than 0.01 per year and for advanced LIGO the rate drops to a peak of $\sim 4$ per year.  This is because the clusters that produce few mergers serve only to dilute the
detection rate.  4 BH-BH detections per year is, however, still twice that predicted by \cite{Belczynski07} for galactic field populations and thus star
clusters still provide a modest enhancement to the BH-BH detection rate.

\section{Detection Rates for Space-Based Detectors}
\label{sec:LISAdet}
Most of the escaper binaries that are potential LISA sources will be in the LISA band for period of time that is less than the spread in ages of Galactic globular clusters. However, if a source is in the LISA band, it will appear as a quasi-monochromatic source for the duration of the LISA mission. Therefore, we will express the detectability of escaper BH binaries in terms of the probability that a source will be in band during the LISA mission. We do this by distributing the population of escaper BH binaries throughout the Galactic halo and determining the lifetime of each detectable source. Using the spread in globular cluster ages, we can then express the probability of detection as a ratio of the lifetime to the spread in globular cluster ages. For this paper, we adopt a spread of 4 Gyr.

\begin{figure}
\centering
\includegraphics[clip,width=0.45\textwidth]{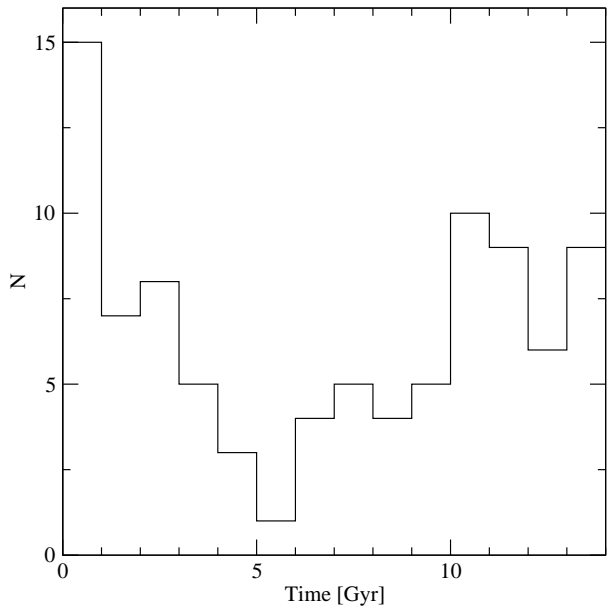}
\caption[Escape Time for LISA]{The total number of escape times for LISA source BH-BH escapers per Gyr summed over all 160 simulations.\label{fig:escapetime}}.
\end{figure}

In order for a binary to be a LISA source, it must lie within the LISA sensitivity band during the expected age range of the Galactic globular clusters. We take the age range to be between 10 and 14 Gyr~\citep{forbes10}. We determine the if a binary is potentially within the LISA band by considering whether the harmonic with maximum power has a frequency above the bottom of LISA's sensitivity band, which we take to be $5\times10^{-5}$ Hz. Out of all 160 simulations, there are 91 escaped binaries within the LISA band during the 10-14 Gyr time-frame. The number of escapers within the LISA band are binned  by Gyr and shown in figure~\ref{fig:escapetime}.

\begin{figure}
\centering
\includegraphics[clip,width=0.45\textwidth]{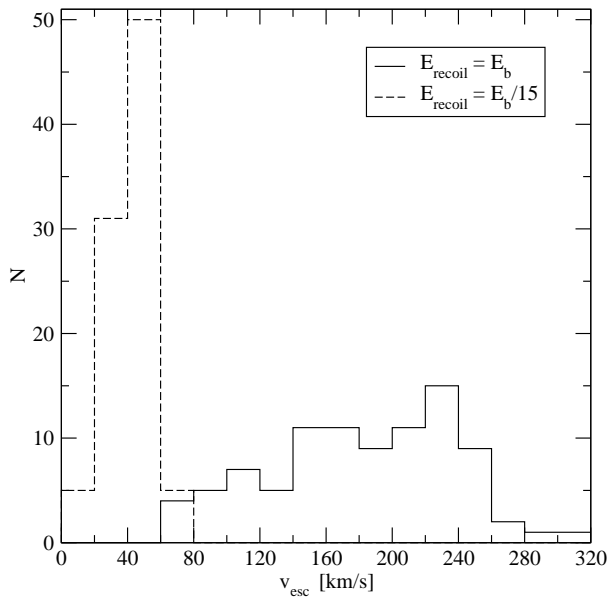}
\caption[Escape Velocity for LISA]{The total number of recoil velocities for LISA source BH-BH escapers per 20 km/s interval. The solid line is the upper bound, assuming 50\% of the initial binding energy goes into recoil energy during the last interaction. The lower bound, using the prescription of Portegies Zwart and McMillan~\cite{SPZandM00} is the dashed line.\label{fig:escapevelocity}}
\end{figure}

\begin{figure}
\centering
\includegraphics[clip,width=0.45\textwidth]{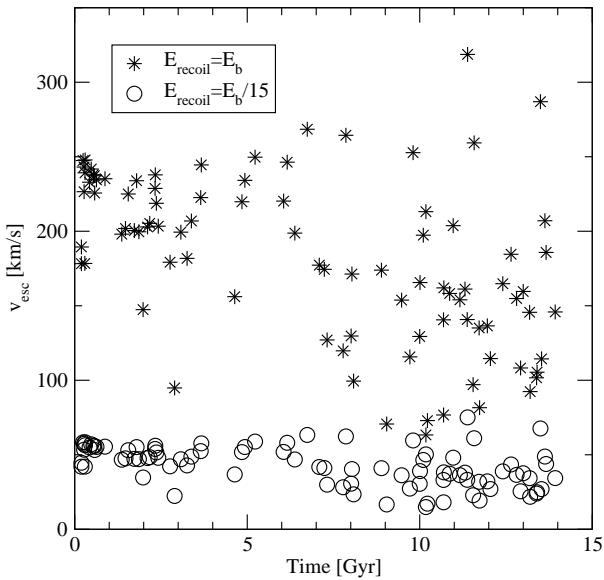}
\caption[Escape Velocity vs Time for LISA]{The estimated escape velocity plotted against the time of escape for both upper and lower bounds.\label{fig:vescvstime}}
\end{figure}

In order to distribute the escapers throughout the Galactic halo, we need to adopt a density distribution for the escaped binaries. Because we have not explicitly modeled individual globular clusters, we do not trace the path of the escaped binaries through the Galactic potential. Instead, we note that the average binding energy of the escapers can give an estimate of their recoil velocity following the last encounter. For an upper bound on the recoil velocity, we assume that the binding energy of the escaped binary is equal to its recoil kinetic energy, while for a lower bound, we follow \cite{SPZandM00} and assume that the recoil energy is $1/15$ of the final binding energy. We show the number binaries binned by $20~{\rm km/s}$ intervals for all escapers within the LISA band in figure~\ref{fig:escapevelocity}, using both prescriptions for the recoil energy. We note also that the escape velocity distribution is fairly flat over the lifetime of the cluster, with a slight trend toward lower velocities at later times (although the spread in velocities increases). This is shown in figure~\ref{fig:vescvstime} for both the upper and lower bounds of velocities. 


The lower bound recoil velocities are comparable to the escape velocity for a typical globular cluster, and consequently the escapers are most likely to be found at Galacto-centric radial distances that are comparable to the radial distribution of the globular cluster system. We use the Harris catalog~\cite{harris96} to obtain these distances. On the other hand, the upper bound recoil velocities are comparable to the Galactic rotational velocities at globular cluster distances. In this case, the escapers are likely to be found distributed according to the Galactic halo distribution. We use a simplified spherical density profile~(\cite{zinn85,morrison96,siegel02}):
\begin{equation}
\rho_{\rm halo} \propto \left(1 + \frac{r}{a_0}\right)^{-3.5}
\end{equation}
where $a_0 = 3.5~{\rm kpc}$ is the scale radius of the halo. We impose a cut-off radius of $r_{\rm co} = 300~{\rm kpc}$

Using a Monte Carlo technique, we place each potentially detectable BH-BH binary in 100 locations determined by the globular cluster distribution for an upper bound on the detection likelihood and 100 locations determined by the halo distribution for the lower bound. The initial phase and orientation are randomly chosen at each location. For each location, we compute the orbital period and eccentricity when the binary first becomes detectable in the LISA data stream. We determine detectability by imposing a detection threshold on the signal-to-noise ratio ($\rho$), computed according to:
\begin{equation}
\rho^s = 4\int_0^\infty{\frac{\left|\tilde{h}(f)\right|^2}{S_n(f)}df}
\end{equation}
where $\tilde{h}(f)$ is the Fourier transform of the response of  LISA to the gravitational wave and $S_n(f)$ is the power spectral density of the expected noise in LISA. We include both instrument noise and an estimate of the Galactic white dwarf binary foreground from~\citet{Ruiter10}. We compute the response of LISA using the time domain as described in~\citet{Benacquista04}, taking the barycentered waveform for eccentric binaries from~\citet{Pierro01}, carried out to the $n\sim 1300$ harmonic. We set a detection threshold of $\rho \ge 10$ in a single interferometer (which corresponds to a combined $\rho \ge 14$ in two channels of the LISA data stream).

\begin{figure}
\centering
\includegraphics[clip,width=0.45\textwidth]{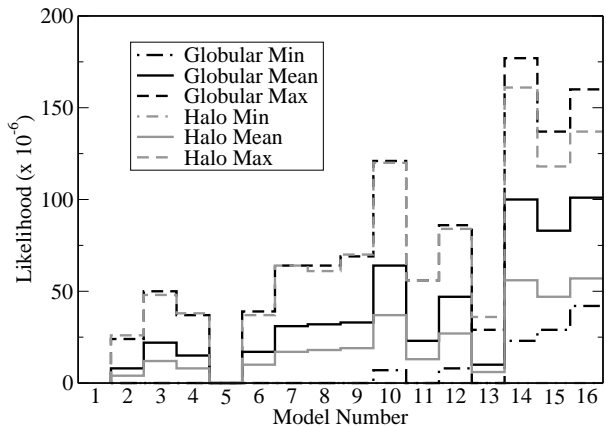}
\caption[Likelihood of Detection]{Likelihood of LISA detection of an escaped BH-BH binary from a single globular cluster for each model. Model numbers correspond to the order in which models are listed in Table~\ref{tab:LISAlifetimes}.\label{fig:LISAlikelihoods}}.
\end{figure}

Once the initial orbital period and eccentricity at the onset of detection have been determined, we compute the lifetime of the binary in the LISA band using~\citet{Peters64} to find the time to coalescence. We note that the actual coalescence will occur above of the LISA band, but that the time spent in this phase is negligible compared with the time spent in the LISA band. Thus, for each BH-BH binary, we have 100 possible lifetimes in the LISA band, which we condense into a mean lifetime and its associated variance. In order to estimate the likelihood of a given model producing a detectable BH-BH binary from its population of escapers, we average over the lifetimes of each BH-BH binary in each model. We also average over the variances to obtain a spread in the lifetimes. We do this for both the globular cluster distribution and the halo distribution. The results are shown in Table~\ref{tab:LISAlifetimes}. We also show likelihoods in Figure~\ref{fig:LISAlikelihoods}. We note that the models with low metallicity and high initial binary fraction have the highest likelihood of ejecting a BH-BH binary that will be detectable with LISA.  However, even in the most optimistic case with every Galactic globular cluster represented by the highest likelihood model (50low37), the likelihood of LISA detection is less than 3 \%. Thus, detection of a BH-BH binary produced in a Milky Way globular cluster is unlikely.

\begin{figure*}
\centering
\includegraphics[width=\textwidth]{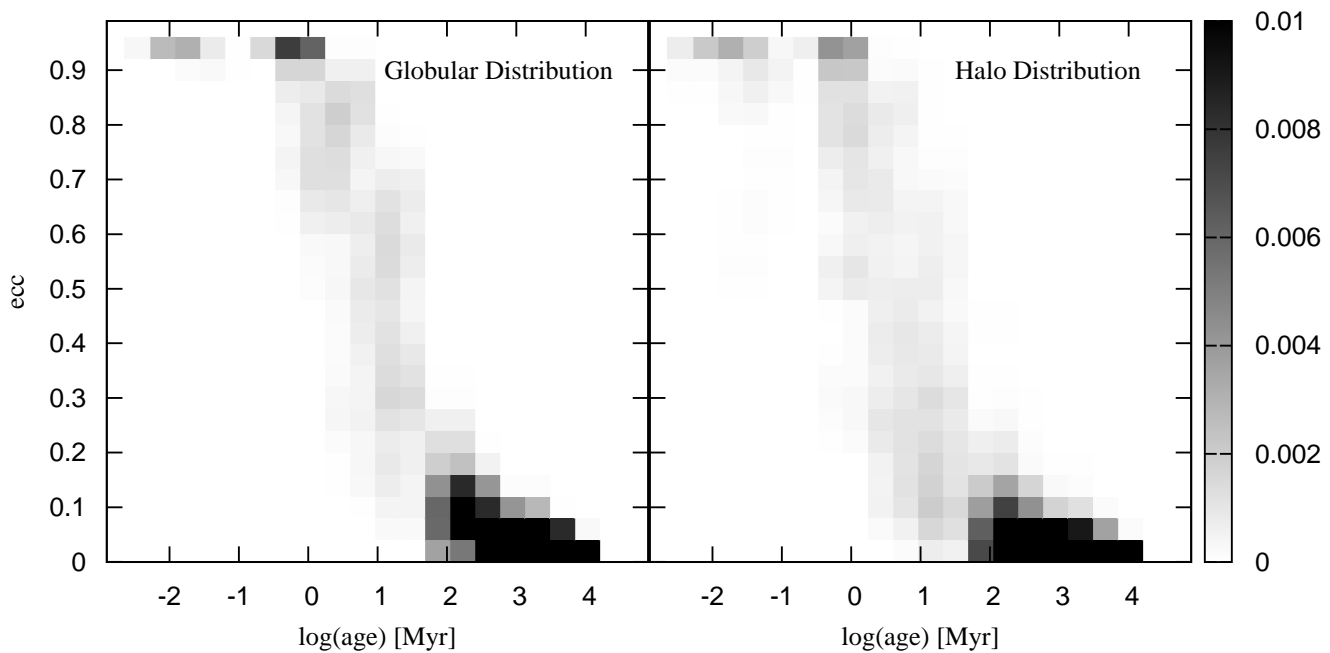}
\caption[LISA Eccentricities]{The eccentricity of BH-BH binaries in the LISA band as a function of the time they have been in the galactic field for both the globular cluster and the halo distributions.\label{fig:LISAecc}}
\end{figure*}

The majority of BH-BH binaries in the LISA band are close to circular with very little model-to-model variation.  There are, however, a few binaries with very high eccentricity and, as is shown in Figure~\ref{fig:LISAecc}, these are the ones that have recently escaped from their parent clusters.  This result is not surprising since the binaries that have been in the field for the shortest period of time are closer to their last interaction and thus to the last chance for signficant eccentricity to be imparted before gravitational wave radiation carries it away.  This also agrees with the results of Paper I where all the LISA binaries that remain in the clusters and are subject to frequent interactions remain highly eccentric.  This opens a small (but non-zero) possibility of a highly eccentric stellar mass detection in the LISA band.

\begin{table}
\centering
\caption[Lifetimes and Detection Probabilities for LISA]{The average lifetime and likelihood for LISA detection for escapers using Globular cluster radial distribution or halo radial distribution. Errors are one standard deviation obtained from 100 realisations of each distribution.\label{tab:LISAlifetimes}}
\scriptsize{
\begin{tabular}[c]{l r r r r}
\hline
&\multicolumn{2}{c}{Globular Cluster}&\multicolumn{2}{c}{Halo}\\
\hline
Simulation & Lifetime & Likelihood & Lifetime & Likelihood \\
& ($\times 10^3$ yr)&($\times 10^{-6}$)&($\times 10^3$ yr)&($\times 10^{-6}$)\\
\hline
10sol21  & \multicolumn{1}{c}{---}   & \multicolumn{1}{c}{---} & \multicolumn{1}{c}{---} & \multicolumn{1}{c}{---} \\
10sol37  &  $32\pm\phantom{1}65$ &  $8\pm \phantom{1}16$ & $18\pm\phantom{1}89$ &  $4\pm\phantom{1}22$ \\
10sol75  &  $86\pm114$ &  $22\pm\phantom{1}28$ & $47\pm143$ &  $12\pm\phantom{1}36$ \\
10sol180 &  $60\pm\phantom{1}89$ &  $15 \pm \phantom{1}22$ &  $33\pm118$ &  $8 \pm \phantom{1}30$ \\
50sol21  & \multicolumn{1}{c}{---}   & \multicolumn{1}{c}{---} & \multicolumn{1}{c}{---} & \multicolumn{1}{c}{---} \\
50sol37  & $70\pm\phantom{1}86$ &  $17\pm\phantom{1}22$ & $38\pm107$ &  $10\pm\phantom{1}27$ \\
50sol75  & $123\pm132$ & $31\pm\phantom{1}33$ & $69\pm187$ &  $17\pm\phantom{1}47$ \\
50sol180 & $126\pm129$ & $32 \pm \phantom{1}32$ &  $71\pm173$ &  $18 \pm \phantom{1}43$ \\
10low21  &  $132 \pm 144$ &  $33 \pm \phantom{1}36$ & $76 \pm 202 $ &  $19 \pm \phantom{1}51$ \\
10low37  &  $257\pm230$ &  $64\pm\phantom{1}57$ & $146\pm331$ &  $37\pm\phantom{1}83$ \\
10low75  &  $93\pm133$ &  $23 \pm \phantom{1}33$ &  $51\pm171$ &  $13 \pm \phantom{1}43$ \\
10low180 & $187\pm154$ & $47 \pm \phantom{1}39$ &  $108\pm226$ &  $27 \pm \phantom{1}57$ \\
50low21  & $40 \pm \phantom{1}77$ & $10 \pm \phantom{1}19$ & $24 \pm 121$ &  $6 \pm \phantom{1}30$ \\
50low37  & $400\pm308$ & $100\pm\phantom{1}77$ &  $225\pm420$ &  $56\pm105$ \\
50low75  & $333\pm214$ & $83 \pm \phantom{1}54$ &  $189\pm284$ &  $47 \pm \phantom{1}71$ \\
50low180 & $405\pm234$ & $101 \pm \phantom{1}59$ & $227\pm319$ & $57 \pm \phantom{1}80$ \\
\hline
\end{tabular}
}
\end{table}

\begin{figure}
\centering
\includegraphics[clip,width=0.45\textwidth]{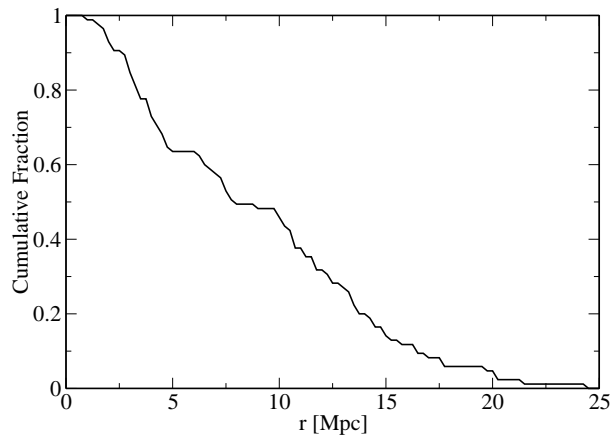}
\caption[Max Distance to Detection]{Cumulative histogram showing the fraction of binaries in the LISA sensitivity band that are detectable at a distance $r$.\label{fig:MaxDistance}}
\end{figure}

Since BH-BH binaries are strong sources of gravitational radiation, they can be observed at distances substantially greater than the scale of the Milky Way. We can estimate the maximum distance to which a BH-BH binary can be detected after an observation time $T$ using:
\begin{equation}
r_{\rm max} = \frac{h_{\rm c}\sqrt{T}}{\rho \sqrt{S_n}},
\end{equation}
where $h_{\rm c}$ is the characteristic strain at a given gravitational wave frequency, $\rho$ is the threshold snr for detection, and $\sqrt{S_n}$ is the strain spectral density at that frequency. For $h_{\rm c}$, we use~\citep{benacquista99}:
\begin{equation}
h_{\rm c} = 2(4\pi)^{1/3}\frac{G^{5/3}{\mathcal{M}_{\rm chirp}}^{5/3}f_{\rm GW}^{2/3}}{rc^4}.
\label{hchar}
\end{equation}
From the {\em LISA Sensitivity Curve Generator} available at {\tt http://www.srl.caltech.edu/\~{}shane/sensitivity/}~\citep{larson02}, we find the minimum strain spectral density to be $\sqrt{S_n} = 1.2\times10^{-20}/\sqrt{\rm Hz}$ at $f_{\rm GW} = 7~{\rm mHz}$. Setting a detection threshold of $\rho=10$, we find that the median maximum distance for all detectable binaries to be $r_{\rm max} \sim 7.5$ Mpc. The cumulative histogram of $r_{\rm max}$ for all detectable binaries in all models is shown in Figure~\ref{fig:MaxDistance}.

\section{Discussion}
\label{sec:discussion}

There are as yet no detections of gravitational waves in either the high-frequency ground-based or low-frequency space-based regimes.  We can, however, compare our ground-based results to previous studies of BH-BH binaries in globular clusters using different dynamical assumptions and to some tentative predictions for stellar-mass LISA sources in the Galactic field.

As previously mentioned, the investigations of \cite{OLeary06} and \cite{Sadowski08} are some of the most complete available and make opposing dynamical assumptions regarding mass-segregation of BHs and thus the interaction rate in the core.  In the work of \cite{OLeary06} BHs are assumed to mass segregate completely and form a strongly interacting sub-system in the core of the cluster.  This leads to a large number of interactions that can both create and destroy BH-BH binaries.  By contrast \cite{Sadowski08} assume that the BHs do not mass-segregate and remain in thermal equilibrium with the rest of the stars in the cluster.  This leads to a much lower interaction rate between the BHs and, consequently, a lower rate of both BH-BH formation and destruction.  In practise the lower destruction rate wins and the simulations of \cite{Sadowski08} predict a yearly detection rate at least an order of magnitude higher than that of \cite{OLeary06}.  This implies that the predictions of \cite{OLeary06} represent a lower limit on the BH-BH merger detection rate while those of \cite{Sadowski08} represent an upper limit.  \cite{OLeary06} also predict that up to 70\% of BH-BH mergers will occur in binaries that have been ejected from the cluster whereas \cite{Sadowski08} predict that $\sim 90$\% of mergers occur in binaries remaining within the clusters.

The predictions of our simulations, using a fully self-consistent treatment of stellar dynamics, agree with those of \cite{OLeary06} better than those of \cite{Sadowski08}.  We confirm that few, if any, mergers should take place within the clusters and the peak detection rates predicted for our denser clusters in Figure~\ref{fig:RatesSingle} fall in the same range as those given in Table 2 of \cite{OLeary06}.  This agreement is in spite of the fact that both \cite{OLeary06} and \cite{Sadowski08} use a similar and much more detailed treatment of few-body interactions than is present in our Monte Carlo code.  For these reasons we favour the approximations made by \cite{OLeary06} rather than those made by \cite{Sadowski08} and postulate that a proper treatment of global globular cluster dynamics may be more important in predicting event \emph{rates} for ground-based detectors than the detailed microphysics of the interactions.

The most significant difference between our predictions and those of \cite{OLeary06} is that we do not find any mergers between BH-BH binaries that remain in our clusters, only in the escapers.  As we demonstrated in Paper I, this is because in our clusters BH-BH binaries tend to either be disrupted or ejected by dynamical interactions before they have a chance to merge due to gravitation wave radiation.  It is possible that this effect is real, as suggested by the results of \cite{SPZandM00}, however it may also be due to our more approximate treatment of few-body interactions.  As discussed in Paper I we may be overestimating both the disruption rate in binary-binary interactions and the velocity kick applied to the centre-of-mass of each member of an interaction while at the same time underestimating the eccentricities of binaries formed through these interactions.  This combination of effects would lead to shorter disruption timescales, larger escape probabilities and longer gravitational wave inspiral timescales than would be expected in a real cluster.  Thus it is almost certain that our merger rates, particularly those within the cluster, are lower limits.  A version of our Monte Carlo code with a more detailed treatment of few-body interactions is currently in development and will allow us to investigate our BH-BH escape rates in more detail.

We have no direct comparisons in the LISA band since all studies of globular clusters as sources of \emph{resolved} stellar-mass LISA sources have focused only on those binaries that are retained by the cluster (e.g.~\cite{Benacquista01,benacquista01, kocis06, ivanova06, willems07}). 

Much more attention has been paid to field binaries. Both~\cite{Nelemans01} and \cite{BBandB08} have investigated the possibility that BH-BH binaries in the Galactic field could be resolved stellar-mass LISA sources.  \cite{BBandB08} further compared two different models for BH-BH binary formation in the Galactic field. They found that if binaries can survive the Hertzsprung gap without merging then at the present time there may be $\sim 10$ resolvable double compact binaries in the Galactic field, half of which may be BH-BH.  By contrast if most binaries merge during the Hertzsprung gap then there would only be $\sim 2$ resolvable double compact binaries in the Galactic field, none of which are expected to be BH-BH.  Taking our estimate that our 4 Gyr ``snapshot'' represents the current Galactic globular cluster system at face value and noting 85 out of our 91 LISA binaries are resolvable we find $\sim 1 \times 10^{-1}$ resolvable BH-BH LISA sources per initial $10^{5}$M$_{\odot}$ in globular clusters as opposed to the $\sim 0-1 \times 10^{-5}$ per initial $10^{5}$M$_{\odot}$ in Galactic stellar mass found by \cite{BBandB08}.  Thus globular clusters do seem to efficiently produce LISA sources.  We note, however, that it is very unlikely that all 85 of the resolvable BH-BH binaries we find will be ``on'' at the same time, hence our low detection probability compared to \cite{BBandB08} who's BH-BH binaries will all be on at the same time by construction.  \cite{BBandB08} concluded that there is a reasonable probability that LISA will resolve Galactic double compact binaries and that the presence of BH-BH detections can be used to constrain the behaviour of binaries during the Hertzsprung gap.  Our results do not contradict the first point, although Galactic globular clusters may not contribute much to the LISA detection rate.  They do, however, imply that even if a BH-BH binary is detected in the LISA band it is not necessarily evidence that binaries survive the Hertzsprung gap phase without merging since the binary could be produced by dynamical interactions in a star cluster.  Our results also indicate that if a highly eccentric BH-BH binary is detected by LISA is likely to have experienced a dynamcial interaction in its recent past and thus is likely to have originated in a star cluster.  Note that the converse is not true, a circluar BH-BH binary in the LISA band may be either a cluster binary that has existed in the field for long enough to cirularise or it may have a galactic origin.

The discovery that star clusters produce BH-BH binaries that are visible to LISA up to Mpc distances is very interesting since it indicates that we may be able to detect such binaries in galaxies other than our own.  For this reason we plan to extend our work to other galaxies in our local group and determine if this will have a significant effect on the probability of LISA detecting a stellar-mass BH-BH binary.

\section{Conclusions}
\label{sec:conclusion}

We have studied the behaviour of the BH population in globular clusters, focusing on BH-BH binaries formed by dynamical interactions but then inspiralling and merging in a galactic field.  We find that globular clusters produce such binaries quite efficiently and should enhance the BH-BH binary detection rate for the next generation of ground-based detectors by a modest factor over the rate expected from isolated stellar evolution in galactic fields.  This enhancement is in line with the predictions of \cite{OLeary06} rather than the more extreme factors predicted by \cite{Sadowski08}.  We find that globular clusters will also produce stellar-mass LISA sources resolvable at distances of several Mpc.  The probability of such a binary being "on" in our Galaxy during the three year LISA mission is, however, rather low.  In future we plan to include few-body integration in our code and see if we get better agreement with \cite{OLeary06} on the number of mergers within clusters.  Given the large distance to which LISA can detect some of our binaries, we are also interested in extending our study to other local group galaxies with simulations chosen to more closely match the real globular cluster population.  We conclude that star clusters can efficiently enrich galactic fields with BH-BH binaries and must be taken into account when estimating detection rates and population characteristics for ground- and space-based gravitational wave detectors.

\section*{Acknowledgements}

J.M.B.D. would like to thank the International Max-Planck Research School for Astronomy and Cosmic Physics at the University of Heidelberg (IMPRS-HD) for
providing funding for his Ph.D.  The simulations have been carried out at the High Performance Computing Centre Stuttgart (HLRS) using the resources of
Baden-W\"u{}rttemberg grid (bwgrid) through the German Astrogrid-D and D-Grid projects.  M.J.B acknowledges the support of NASA Grant NNX08AB74G and the
Center for Gravitational Wave Astronomy, supported by NSF award \#{}0734800 and NASA award NNX09AV06A.  M.G. was supported by Polish Ministry of Science and Higher Education through
the grant 92/N.ASTROSIM/2008/0 and N N203 380036.  R.S. thanks the Deutsches Zentrum f\"u{}r Luft- und Raumfahrt (DLR) four support within the LISA Germany project.

\label{lastpage}

\end{document}